\documentclass[onecolumn,superscriptaddress,nofootinbib]{revtex4}

\pdfoutput=1
\usepackage{amssymb,amsmath,amsthm,graphicx}
\usepackage{graphics, color}
\usepackage{subfigure} 
\usepackage{latexsym}
\usepackage{bm}
\usepackage{epsfig}
\usepackage[pdftex]{hyperref}

\begin{document}
\title{Localised and nonuniform thermal states of super-Yang-Mills on a circle}
\author{\'Oscar J.~C.~Dias}
\email{ojcd1r13@soton.ac.uk}
\affiliation{STAG research centre and Mathematical Sciences, University of Southampton, UK}
\author{Jorge E.~Santos}
\email{J.E.Santos@damtp.cam.ac.uk}
\affiliation{DAMTP, Centre for Mathematical Sciences, University of Cambridge, \\ Wilberforce Road, Cambridge CB3 0WA, UK}
\author{Benson Way}
\email{benson@phas.ubc.ca}
\affiliation{Department of Physics and Astronomy, University of British Columbia, 6224 Agricultural Road, Vancouver, B.C., V6T 1W9, Canada}
\begin{abstract}\noindent{
At low energies or temperatures, maximally supersymmetric Yang-Mills theory on $\mathbb R^{(t)}\times S^1$ with large $N$ gauge group $SU(N)$ and strong t'Hooft coupling is conjectured to be dual to the low energy dynamics of a collection of D0-branes on a circle.  We construct thermal states in the gravitational side of the correspondence where we find a first-order phase transition between states that are uniform on the $S^1$ and states that are localised on it.  When compared with lattice computations that are now available, these critical values provide the first instance where a first-order phase transition is tested on both sides of gauge/gravity duality. 
}
\end{abstract}
\maketitle

\tableofcontents
\newpage

\section{Introduction and Review}

\subsection{Introduction}
\noindent\indent Almost two decades ago, a number of gravitational theories were conjectured to be dual to specific strongly coupled large $N$ maximally supersymmetric Yang-Mills (SYM) theories \cite{Mal97,Wit98a,GubKle98,Itzhaki:1998dd}. These conjectures paved the way to our current understanding of the gauge/gravity duality paradigm. However, a proof remains elusive.  The main difficulty lies in the fact that the gravitational descriptions are only valid when the corresponding gauge theory duals are strongly coupled. 

While there is accumulating evidence that these dualities hold, most nontrivial tests are restricted to integrable sectors of the theory \cite{Gromov:2009zb} or to situations with unbroken supersymmetry \cite{Aharony:1999ti,Benini:2016rke} (\emph{i.e.} zero temperature). However, over the last ten years, a number of novel techniques have been used to analyse these field theories at strong coupling using computer lattice simulations \cite{Hanada:2007ti,Catterall:2007fp,Anagnostopoulos:2007fw,Catterall:2008yz,Hanada:2008gy,Hanada:2008ez,Catterall:2009xn,Hanada:2009ne,Hanada:2013rga}. These methods allow nonintegrable sectors of the theory to be tested at finite temperature and provide some of the most nontrivial tests of the gauge/gravity correspondence to date.  Despite these efforts, many important gravitational phenomena have not been directly verified in the corresponding field theory. These include thermodynamic phase transitions like the Hawking-Page transition \cite{Hawking:1982dh,Witten:1998qj,Sundborg:1999ue,Aharony:2003sx} dual to confinement/deconfinement and those arising from a Gregory-Laflamme instability \cite{Gregory:1993vy,Gregory:1994bj,Gregory:1994tw,Susskind:1997dr,Barbon:1998cr,Li:1998jy,Fidkowski:2004fc,Martinec:strings98,Dias:2015pda,Dias:2016eto} dual to symmetry breaking phase transitions. 

Interestingly, several authors have performed lattice simulations of SYM on $\mathbb R^{(t)}\times S^1$ at finite temperature and at strong coupling \cite{Aharony:2004ig,Azeyanagi:2009zf,Hanada:2009hq,Hanada:2010kt,Hanada:2010gs,Catterall:2010fx,Hanada:2016qbz}. In a certain limit, this theory has a conjectured gravity dual, but the corresponding gravitational solutions are unavailable. These results offer the first quantitative predictions for classical gravity coming from the gauge-theory side of the correspondence.  The aim of this work is to complete this test of gauge/gravity by numerically constructing the dual black hole solutions. 
Together with lattice simulations, these results would comprise the first quantitative test of gauge/gravity duality with a first-order phase transition. 

At strong coupling and large $N$ gauge group, the gravity dual to SYM on $\mathbb R^{(t)}\times S^1$ depends upon the temperature (or energy) scale relative to the circle size and dimensionless coupling.  At high temperatures, the gravity description lies within type IIB supergravity, while at low temperatures the dual lies in type IIA supergravity. Within the type IIA regime, the gravity theory exhibits a Gregory-Laflamme instability \cite{Gregory:1993vy,Gregory:1994bj} when the horizon radius of p-branes becomes much smaller than the circle size \cite{Susskind:1997dr,Barbon:1998cr,Li:1998jy,Aharony:2004ig,Aharony:2005ew,Catterall:2010fx}.  The instability implies the existence of time-independent solutions that break the symmetries of the $S^1$.  Such solutions compete with solutions that preserve the $S^1$ symmetry, allowing for the possibility of phase transitions. 

Locating phase transitions on the gravity side of the correspondence requires constructing the competing type IIA supergravity solutions and comparing their thermodynamic quantities.  Fortunately, there is a procedure that generates solutions to type IIA supergravity from solutions of the simpler 10-dimensional vacuum Einstein equation $R_{ab}=0$ with $\mathbb R^{(1,8)}\times S^1$ asymptotics.  The process involves an uplift to 11 dimensions, followed by a boost in the extra direction, then finally a Kaluza-Klein (KK) reduction back to 10 dimensions. Systems with $\mathbb R^{(1,d-2)}\times S^1$ asymptotics are well-studied \cite{Kol:2002xz,Wiseman:2002zc,Kol:2003ja,Harmark:2003yz,Gorbonos:2004uc,Harmark:2004ws,Asnin:2006ip,Harmark:2002tr,Wiseman:2002ti,Kudoh:2003ki,Kudoh:2004hs,Sorkin:2006wp,Kleihaus:2006ee,Harmark:2007md,Dias:2007hg,Headrick:2009pv,Wiseman:2011by,Figueras:2012xj,HorowitzBook2012,Kalisch:2016fkm} and can be used to infer the behaviour of type IIA solutions.  These results suggest that the $d=10$ case of interest contains three types of solutions, corresponding to a uniform phase, a nonuniform phase, and a fully localised phase.  Furthermore, there should be a first-order phase transition from the uniform phase to the localised phase.  However, the only available thermodynamic quantities in $d=10$ correspond to  the uniform phase, which is known analytically.\footnote{The nonuniform phase has been constructed for $5\leq d\leq 15$ \cite{Wiseman:2002zc,Kleihaus:2006ee,Sorkin:2006wp,Figueras:2012xj,Kalisch:2016fkm}. However, as far as we are aware, the thermodynamics of this phase has never been displayed for $d=10$. The localised phase has only been constructed in $d=5,6$ \cite{Sorkin:2003ka,Kudoh:2004hs,Headrick:2009pv}.  Perturbative results are available for the localised phase, which we will describe in Section \ref{ThermoUnifPert}. However, the expected phase transition is located in a non-perturbative regime.}  We will therefore construct the desired nonuniform and localised phases and complete the SYM phase diagram from the gravitational side of the duality. 

Performing lattice simulations on the gauge theory side, the authors of \cite{Catterall:2010fx} find a first-order phase transition at a temperature near the critical zero-mode temperature of the Gregory-Laflamme instability. Our gravitational results indeed locate a first-order phase transition at a temperature ratio of 1.093 with respect to the zero-mode temperature. While our results are consistent with \cite{Catterall:2010fx}, a more precise test will await improved lattice computations. 

This manuscript is structured as follows. In the next two subsections of this Introduction, we describe the broad context of our study. Namely, in subsection \ref{decoupling}, we review D$p$-branes, their decoupling limit, and the dual SYM theory. In subsection \ref{SYMoncircle}, we review the uniform supergravity solutions that are dual $(1+1)-$dimensional SYM on a circle.  Section \ref{MapGRtoSYM} revisits the uplift-boost-KK reduction generating technique that allows us to find solutions of type IIA supergravity by solving the simpler vacuum Einstein equation in 10 dimensions. In particular, we describe a map between the thermodynamics of vacuum gravity and those of $(1+1)-$dimensional SYM on a circle. Section \ref{sec:NumConstruction} describes our numerical construction of the nonuniform and localised solutions, and their corresponding SYM phase diagram is discussed in Section \ref{sec:Results}. We close with a few remarks in \ref{sec:Conclusion}.  For completeness, the equations of motion for the sector of type II supergravity relevant to our study are given in Appendix \ref{sec:EOM}, and the thermodynamics of our solutions within vacuum gravity are displayed in Appendix \ref{sec:ResultsGR}. These results complement similar results available in other dimensions. 

\subsection{Review of D$p$-branes, their decoupling limit, and the dual SYM theory \label{decoupling}}

Let us review the duality between  $(1+1)$ dimensional SYM theory and supergravity \cite{Itzhaki:1998dd}.  We begin by considering $N$ coincident D$p$-branes in type II string theory (we will later restrict to the case $p=1$ in Section \ref{SYMoncircle}).  In general, there are modes that propagate along the worldvolume of the brane and modes that propagate in the bulk.  Ref. \cite{Itzhaki:1998dd} identified a limit where the bulk modes decouple and the worldvolume theory reduces to SYM.  By identifying the low energy sector of the Dirac-Born-Infeld action (describing open string excitations on  D$p$-branes) with the SYM action, one can relate the SYM coupling constant $g_{\mathrm YM}$ to the string length $\ell_s=\sqrt{\alpha^\prime}$ and string coupling $g_s$ via
\begin{equation}\label{gYM}
g_{\rm YM}^2\equiv (2 \pi )^{p-2} g_s \ell_s^{p-3}\,.
\end{equation}
The decoupling limit of \cite{Itzhaki:1998dd}, valid in the t'Hooft large $N$ limit and for strong t'Hooft coupling $\lambda=g_{\rm YM}^2 N$, sends $\ell_s\to0$ while keeping $g_{\rm YM}$ fixed.  This limit suppresses higher-order $\alpha^\prime$ corrections and, for $p\leq3$ (the case relevant here), sends the gravitational Newton's constant $G_{10}$ to zero. Indeed, recall that identifying the low-energy action of type II (closed) string theory  and the action of type II supergravity yields $16 \pi  G_{10}\equiv (2 \pi )^7 g_s^2 \ell_s^8$. This limit can be taken at finite energy $U$ and charge $K_p$, where these quantities must also be held fixed. 
At finite energies $U$, the effective dimensionless SYM coupling is given by $g_{\mathrm{eff}}^2\approx g_{\mathrm{YM}}^2NU^{p-3}$. For $p<3$, perturbative SYM is valid for large $U$, and the theory is UV free. 

On the other hand, a stack of $N$ coincident D$p$-branes can also be described within classical type II supergravity, provided that curvature scales remain small compared to the string scale (to suppress $\alpha'$ corrections) and the effective dimensionless string coupling is sufficiently small (to suppress string loop effects).  This classical theory contains a graviton $g$, dilaton $\phi$ and a Ramond-Ramond (RR) $A_{(p+1)}$ field with action (here in the string frame)
\begin{equation}
\label{action}
I_{II}^{(s)} = \frac{1}{(2 \pi )^7 \ell_s^8} \int \mathrm d^{10} x \sqrt{-g} \Big[ e^{-2\phi } \Big(R
+4 \partial_\mu \phi \partial^\mu \phi \Big) - \frac{1}{2 (p+2)!}  (\mathrm dA_{(p+1)})^2 \Big].
\end{equation} 
For completeness,  the corresponding equations of motion \eqref{IIeomStringFr} and their map to the Einstein frame are given in Appendix \ref{sec:EOM}.
A stack of $N$ coincident non-extremal D$p$-branes (at large $N$) is described within this theory by p-branes (see e.g. \cite{Itzhaki:1998dd,Harmark:1999xt,Harmark:2005jk}):
\begin{eqnarray}\label{nonExtDp}
&& {\mathrm ds}^2= H^{-\frac{1}{2}}\left(-f {\mathrm d}t^{\,2}+{\mathrm d}x_\parallel^2\right)+H^{\frac{1}{2}}\left(\frac{{\mathrm d}r^2}{f}+r^2 {\mathrm d}\Omega_{(8-p)}^2\right),  \nonumber \\
&& e^{\phi}=g_s H^{\frac{3-p}{4}},\qquad \qquad \qquad 
 A_{(p+1)}=(-1)^p g_s^{-1}\coth \beta \left(H^{-1}-1\right) {\mathrm d}t\wedge {\mathrm d}x_1\wedge \cdots \wedge {\mathrm d}x_p,\nonumber \\
&& \hbox{where} \quad f=1-\frac{r_0^{7-p}}{r^{7-p}}, \qquad \quad  H=1+\frac{r_0^{7-p}}{r^{7-p}}\sinh^2\beta,  \qquad \quad {\mathrm d}x_\parallel^2=\sum _{i=1}^p {\mathrm d}x_i^2,
\end{eqnarray}
where the dimensionless string coupling is given by $g_s=e^{\phi_\infty}$, $r_0$ is the horizon location, and $\beta$ is a parameter that sources the gauge potential.
The mass, charge,  temperature, entropy, and chemical potential of these solutions are, respectively,
\begin{eqnarray}
\label{nonExtD$p$Thermo}
&& M_p = V_p \frac{\Omega_{(8-p)}}{16\pi G_{10}} r_0^{7-p}
\Big[ 8-p + (7-p)\sinh^2 \beta \Big], \nonumber \\
&&  Q_p =V_p \frac{\Omega_{(8-p)}}{16\pi G_{10}} (7-p) r_0^{7-p}  \sinh \beta \cosh \beta,
\nonumber \\
&& T_p = \frac{7-p }{4 \pi r_0 \cosh \beta},\nonumber\\
&& S_p = V_p \frac{\Omega_{(8-p)}}{4G_{10}} r_0^{8-p}  \cosh \beta,\nonumber\\
&& \mu_p = \tanh \beta,
\end{eqnarray}
where $\Omega_{n}=\frac{2 \pi ^{\frac{n+1}{2}}}{\Gamma \left(\frac{n+1}{2}\right)}$ is the area of a unit radius S$^n$, $V_p$ is the D$p$-brane worldvolume. These quantities satisfy the thermodynamic first law $\mathrm dM_p = T_p \mathrm dS_p +\mu_p \mathrm dQ_p$ and the Smarr relation $(7-p)M_p=(8-p)T_p S_p+(7-p)\mu_p Q_p$.

One arrives at the conjectured duality \cite{Itzhaki:1998dd} between a gravitational theory and SYM by taking the corresponding decoupling limit of this configuration in classical type II supergravity. To do so, we must complete the relationships between quantities in \eqref{nonExtDp} and stringy quantities $g_s$ and $\ell_s$, and then match them to SYM quantities $g_{\mathrm YM}$, $U$, and $K_p$. The coupling constants are already related via \eqref{gYM} and $16 \pi  G_{10}\equiv (2 \pi )^7 g_s^2 \ell_s^8$.  The relation between charges is given by 
\begin{equation}\label{defKp}
K_p\equiv \frac{(2 \pi )^7 g_s^2 Q_p}{(7-p) V_p \Omega_{(8-p)} \ell _s^{2 (1-p)}}= \frac{(2 \pi)^{7-p} g_s N}{(7-p) \,\Omega_{(8-p)}\ell_s^{3-p}}\;,
\end{equation}
which can be obtained by matching the p-brane charge, computed in the string frame via
\begin{equation}\label{defQp}
Q_p=\frac{V_p}{(2\pi)^7 \ell_s^8}\int_{S^{8-p}}\star \mathrm{d}A_{(p+1)}\,,
\end{equation}
with the charge of $N$ D$p$-branes $Q_p\equiv N \,\mathfrak{t}_p V_p$, where $\mathfrak{t}_p=(2\pi)^{-p}\ell_s^{-(p+1)}g_s^{-1}$ is the D$p$-brane tension.  The energy $U$ can be related via $U=\frac{r}{\ell_s^2}$ and $U_0=\frac{r_0}{\ell_s^2}$. One way to see this is to place one of the D$p$-branes at a position $r$. This configuration breaks the symmetries $U(N)\to U(N-1)\times U(1)$, giving an expectation value (with dimensions of energy) to some of the fields that scales as $r$ \cite{Itzhaki:1998dd}. To summarise, the decoupling limit is given by \cite{Itzhaki:1998dd} 
\begin{equation}\label{NHlim}
 \ell_s\to 0, \qquad g_{\rm YM}^2={\rm fixed}, \qquad {U}\equiv \frac{r}{\ell_s^2}={\rm fixed}, \qquad {U}_0\equiv \frac{r_0}{\ell_s^2}={\rm fixed}, \qquad 
 K_p ={\rm fixed}.
\end{equation}
The result of applying this decoupling limit to \eqref{nonExtDp} gives the near-horizon limit of $N$ non-extremal D$p$-branes\footnote{Note that we gauge away the constant term in $A_{(p+1)}$ and used the fact that $\beta\to\infty$ in the decoupling limit. Note that $\beta$ itself can be related to SYM quantities in this limit via \eqref{defKp}.}:
\begin{eqnarray}\label{NHnonExtDp}  
&& {\mathrm ds}^2= \ell_s^2\left[ \frac{U^{\frac{7-p}{2}}}{\sqrt{d_p} g_{\rm YM} \sqrt{N}}\left(-\left( 1-\frac{U_0^{7-p}}{U^{7-p}} \right) {\mathrm d}t^2+{\mathrm d}x_\parallel^2\right)+ \frac{\sqrt{d_p} g_{\rm YM} \sqrt{N}}{U^{\frac{7-p}{2}}}\left(\left( 1-\frac{U_0^{7-p}}{U^{7-p}} \right)^{-1} {\mathrm d}U^2+U^2 {\mathrm d}\Omega_{(8-p)}^2\right) \right],  \nonumber \\
&& e^{\phi}=(2\pi)^{2-p}\,g_{\rm YM}^2\, \frac{\left(d_p g_{\rm YM}^2 N \right)^{\frac{1}{4}(3-p)}}{U^{\frac{1}{4}(3-p)(7-p)}} ,\qquad \qquad  
A_{(p+1)}=(-1)^p (2\pi)^{p-2}\ell_s^{p+1}\, \frac{U^{7-p}}{d_p g_{\rm YM}^4 N}  {\mathrm d}t\wedge {\mathrm d}x_1\wedge \cdots \wedge {\mathrm d}x_p\,,
\nonumber \\
&& \hbox{with} \:\: d_p\equiv 2^{7-2 p} \pi ^{\frac{3 (3-p)}{2}} \Gamma \left(\frac{7-p}{2}\right).
\end{eqnarray}

The energy above extremality, $\mathcal{E}_p\equiv M_p-Q_p$, the entropy $\mathcal{S}_p$ and temperature $\mathcal{T}_p$ of these near-horizon solutions are 
\begin{eqnarray}
\label{NHnonExtD$p$Thermo}
\mathcal{E}_p = V_p\frac{p-9}{8 \pi ^2 (p-7)} \frac{1}{d_p}   \frac{1}{ g_{\rm YM}^4}  U_0^{7-p},\qquad\qquad 
\mathcal{S}_p =V_p \frac{1}{\pi  (7-p)} \frac{\sqrt{N}}{\sqrt{d_p}} \frac{1}{g_{\rm YM}^3}U_0^{\frac{9-p}{2}}\qquad 
\mathcal{T}_p = \frac{7-p}{4 \pi } \frac{1}{\sqrt{d_p N}} \frac{1}{g_{\rm YM}}U_0^{\frac{5-p}{2}}.
\end{eqnarray}
which obey the thermodynamic first law $\mathrm d\mathcal{E}_p = \mathcal{T}_p \mathrm d\mathcal{S}_p $ and the Smarr relation $\mathcal{E}_p=\frac{1}{2}\frac{9-p}{7-p} \,\mathcal{T}_p \mathcal{S}_p$.

As we have mentioned, classical supergravity is only valid when curvature scales are much smaller than the string scale and for small dimensionless string coupling. The curvature is given by $\alpha^\prime R\sim1/{g_{\mathrm{eff}}}$, and the effective string coupling goes as $e^\phi\sim g_{\mathrm{eff}}^{(7-p)/2}/N$.  Together, these give \cite{Itzhaki:1998dd}
\begin{equation}\label{sugravalidity}
1\ll g_{\mathrm{eff}}^2\ll N^{4/(7-p)}\;.
\end{equation}
Thus, the validity of classical supergravity requires the dual SYM theory to be strongly coupled and $N$ to be large. Note that for $p\neq3$, $g_{\mathrm{eff}}$ depends on the energy $U$, so at fixed $g_{\mathrm{YM}}$, the validity of the supergravity theory depends upon the energy (or temperature). For future use, it will be convenient to rewrite the validity window \eqref{sugravalidity} in terms of the temperature for $p\neq 3$:
\begin{equation}\label{sugravalidityT}
\lambda ^{\frac{1}{3-p}} N^{-\frac{2 (5-p)}{(7-p) (3-p)}}\ll  \mathcal T_p \ll \lambda ^{\frac{1}{3-p}}\,,
\end{equation}
where we have introduced the t'Hooft coupling $\lambda\equiv g_{YM}^2 N$.
Note that the lower bound in  \eqref{sugravalidityT} ensures that the dilaton (at the horizon) is small and thus string coupling corrections are suppressed (i.e. it corresponds to the upper bound in \eqref{sugravalidity}).
On the other hand, the upper bound in \eqref{sugravalidityT} is required to have small curvature in string units (so that $\alpha'$ corrections are negligible), i.e. it corresponds the lower bound in \eqref{sugravalidity}. 
\subsection{Review of supergravity duals to SYM on a circle
\label{SYMoncircle}}

We now restrict ourselves to the $p=1$ case of interest where type II supergravity theory is more specifically type IIB. Moreover, here henceforth we will drop the index $p=1$ from all the expressions, to avoid the proliferation of indices.  Additionally, we take the brane direction to be compactified on a circle $S^1$ with length $L$.  In this case, the gauge theory is (1+1)-dimensional SYM on $\mathbb R^{(t)}\times S^1$ with dimensionless t'Hooft coupling 
\begin{equation}\label{lambdaprimedef}
\lambda^{\prime}\equiv g_{\rm YM}^2 N L^2.
\end{equation}
To describe the gravity side, we take ${\mathrm d}x_\parallel\equiv\mathrm dx$ with identification $x \sim x+L$ and introduce the dimensionless coordinates, $\theta=\frac{2\pi}{L}\,x, u= L \,U, u_0=L \,U_0$. We now have $\theta \sim \theta+2\pi$ and \eqref{NHnonExtDp} reduces to 
\begin{eqnarray}\label{NHnonExtD1}   
&& {\mathrm ds}^2= \ell_s^2\left[ \frac{u^3}{\sqrt{d_1 \lambda^{\prime}}} \left(-\left( 1-\frac{u_0^6}{u^6} \right) \frac{{\mathrm d}t^2}{L^2}+ \frac{{\mathrm d}\theta^2}{(2 \pi )^2} \right)+\frac{\sqrt{d_1 \lambda^{\prime}}}{u^3} \left(\left( 1-\frac{u_0^6}{u^6} \right)^{-1} {\mathrm d}u^2+u^2 {\mathrm d}\Omega_{(7)}^2\right) \right],  \nonumber \\
&&  e^{\phi}=2 \pi \frac{\lambda^{\prime} }{N} \frac{\sqrt{d_1 \lambda^{\prime}}}{u^3},
 \qquad  \qquad   A_{(2)}=- \frac{\ell_s^2}{(2\pi)^2 L}\,\frac{N}{\lambda^{\prime}} \,\frac{u^6}{d_1\lambda^{\prime}}\,{\mathrm d}t \wedge {\mathrm d}\theta, \qquad  \hbox{with} \quad d_1=2^6 \pi ^3.
\end{eqnarray}
This geometry has a horizon with topology $S^7\times S^1$.  Its dimensionless energy above extremality $\varepsilon \equiv\mathcal{E} L$, entropy $\sigma\equiv\mathcal{S}$, and temperature $\tau\equiv \mathcal{T}L$ are 
\begin{equation}
\label{ThermoNHD1}
\varepsilon = \frac{1}{384 \pi ^5} \frac{N^2}{ \lambda^{\prime\,2}}\,u_0^6, \qquad  \qquad  \sigma =\frac{1}{48 \pi ^{5/2}}\frac{N^2}{ \lambda^{\prime\,\frac{3}{2}}}\,u_0^4, \qquad  \qquad  \tau =\frac{3}{16 \pi ^{5/2}}  \frac{u_0^2}{ \sqrt{\lambda^{\prime}} }.
\end{equation}
Provided that the supergravity limit remains valid, it is conjectured \cite{Itzhaki:1998dd,Aharony:2004ig} that \eqref{NHnonExtD1}-\eqref{ThermoNHD1} is the gravitational dual to the uniform phase at temperature $\tau$ of (1+1)-dimensional SYM on the circle $S^1$, in the 't Hooft large $N$ limit and large t'Hooft coupling $\lambda^{\prime}$. Accordingly, the uniform SYM thermal state holographically dual to  \eqref{NHnonExtD1} has energy $\varepsilon$, entropy $\sigma$ and temperature $\tau$ given by \eqref{ThermoNHD1}.  

Now let us discuss the validity of the supergravity description \eqref{NHnonExtD1}  in terms of the temperature $\tau$ \cite{Itzhaki:1998dd,Aharony:2004ig}. The requirements \eqref{sugravalidityT}  become $\sqrt{\lambda^\prime}/N^{2/3}\ll\tau\ll\sqrt{\lambda^\prime}$.  When the lower bound is crossed, the effective string coupling becomes large, but one can still obtain a supergravity description via S-duality \cite{Itzhaki:1998dd}.  For our purposes, we will not take the S-dual and instead take $N$ to be sufficiently large so that this bound is not crossed.

However, there are now two additional requirements which comes from the circle compactification.   First, the curvature scale must be small enough so that excitations of string winding modes wrapping the $S^1$ are suppressed. In the neighbourhood of the horizon, the mass of winding modes in string units ($M_w \ell_s$) is given by the winding number $L_H/\ell_s$ with $L_H=\sqrt{g_{\theta\theta}}{\bigl |}_H$. The supergravity description \eqref{NHnonExtD1} is valid when these winding modes are massive in curvature scale units,
$M_w/\sqrt{R}\sim \lambda^{\prime\,\frac{1}{4}} \sqrt{ \tau}\gg 1$
, which implies $\tau\gg 1/\sqrt{\lambda^\prime}$.  Second, perturbations that carry momentum along the circle must not excite string oscillators. This requires that  the length of the circle near the horizon is large in string units, or $L_H/\ell_s\gg 1$, which implies $\tau \gg \lambda^{\prime\,-1/6}$.

Note that together, the upper bound $\tau\ll \sqrt{\lambda^\prime}$ and lower bound  $\tau \gg \lambda^{\prime\,-1/6}$ (or also $\tau\gg 1/\sqrt{\lambda^\prime}$) imply that $\lambda^\prime\gg1$. This in turn picks out one of the lower bounds as more restrictive.  Altogether, we have that the supergravity description \eqref{NHnonExtD1} is good when
\begin{equation}\label{D1valid}
\sqrt{\lambda^\prime}/N^{2/3}\ll\tau\,,\qquad  \lambda^{\prime\,-1/6} \ll  \tau \ll \sqrt{\lambda^{\prime}}\,,\qquad (\lambda^{\prime}\gg 1)\;,
\end{equation}
This supergravity approximation breaks down for small temperatures $ \tau \lesssim \lambda^{\prime\,-1/6}$. Since $\lambda^{\prime}=g_{\rm YM}^2 N L^2$, this critical temperature can be considerably high for a small circle length $L$.

As pointed out in  \cite{Itzhaki:1998dd,Aharony:2004ig}, for temperatures below $\lambda^{\prime\,-1/6}$, one should T-dualise in the $\theta$-direction to obtain a new valid supergravity description. Indeed, T-duality transforms the circle length and string coupling as 
\begin{equation}\label{Tduality}
L\to \widetilde{L}=(2 \pi )^2 \ell _s^2\frac{1}{L},\qquad  g_s \to \widetilde{g}_s =\frac{2\pi \ell_s}{L} g_s.
\end{equation}
Let $z$ be a coordinate with dimension of length such that $z\sim z+2\pi \ell_s$. T-duality transforms this coordinate into a coordinate $\widetilde{z}$ still with dimension of length and with identification $\widetilde{z}\sim \widetilde{z} +2\pi \ell_s$.  The associated  T-dual Buscher transformation rules \cite{Buscher:1987sk,Buscher:1987qj,Lunin:2001fv} for the NS fields are
\begin{eqnarray}
&&e^{2\widetilde{\phi}}=\frac{e^{2\phi}}{g_{zz}},\qquad \widetilde{g}_{zz}=\frac{1}{g_{zz}},\qquad  \widetilde{g}_{\mu z}=\frac{B_{\mu z}}{g_{zz}},\qquad  \widetilde{g}_{\mu \nu}=g_{\mu \nu}-\frac{g_{\mu z}g_{\nu z}-B_{\mu z}B_{\nu z}}{g_{zz}}, \nonumber\\
&&\widetilde{B}_{\mu z}=\frac{g_{\mu z}}{g_{zz}}, \qquad
\widetilde{B}_{\mu \nu}=B_{\mu \nu}-\frac{B_{\mu z}g_{\nu z}-g_{\mu z}B_{\nu z}}{g_{zz}},
\end{eqnarray}
where we note that the NS 2-form $B$ is absent in our solution and these rules are valid in the string frame.  For the RR potentials $A_{(p)}$ the rules are:
\begin{eqnarray}\label{RRTDual1}
{\widetilde{A}}^{(p)}_{\mu\dots\nu\alpha z}&=&A^{(p-1)}_{\mu\dots\nu\alpha}-
(p-1)\frac{A^{(p-1)}_{[\mu\dots\nu| z}\, g_{|\alpha]z}}{g_{zz}},\nonumber\\
{\widetilde{A}}^{(p)}_{\mu\dots\nu\alpha\beta} 
&=&
A^{(p+1)}_{\mu\dots\nu\alpha\beta z}
+pA^{(p-1)}_{[\mu\dots\nu\alpha}\,g_{\beta]z}
+p(p-1)\frac{A^{(p-1)}_{[\mu\dots\nu| z}\,B_{|\alpha |z}\,g_{|\beta ]z}}{g_{zz}}.
\end{eqnarray}
Applying these Buscher rules, with the identifications $z\equiv \ell_s \theta$, $\widetilde{z}\equiv \ell_s \widetilde{\theta}$ and $p=1$, to \eqref{NHnonExtD1}  one gets \footnote{It is permissible to T-dualise the near-horizon geometry \eqref{NHnonExtD1} rather than the full geometry since taking the near-horizon limit commutes with T-duality.} 
\begin{eqnarray}\label{NHnonExtD0}  
&& {\mathrm d\widetilde{s}}^{\,2}= \ell_s^2\left[- \frac{u^3}{\sqrt{d_1 \lambda^{\prime}}} \left( 1-\frac{u_0^6}{u^6} \right) \frac{{\mathrm d}t^2}{L^2} +\frac{\sqrt{d_1 \lambda^{\prime}}}{u^3} \left(\left( 1-\frac{u_0^6}{u^6} \right)^{-1} {\mathrm d}u^2 +u^2 {\mathrm d}\Omega_{(7)}^2 + (2 \pi )^2 {\mathrm d}\widetilde{\theta}^{\,2} \right) \right],  \nonumber \\
&&  e^{\widetilde{\phi}}=(2 \pi)^2 \frac{\lambda^{\prime} }{N} \left(\frac{d_1 \lambda^{\prime}}{u^6}\right)^{3/4},
 \qquad  \qquad   \widetilde{A}_{(1)}=-\frac{1}{(2\pi)^2} \frac{\ell_s}{L} \,\frac{N}{\lambda^{\prime}} \,\frac{u^6}{d_1\lambda^{\prime}}\,{\mathrm d}t, \qquad  \hbox{with} \quad d_1=2^6 \pi ^3,
\end{eqnarray}
where $ \widetilde{\theta}\sim \widetilde{\theta}+2\pi$. 
After the T-duality, the solution of type IIB supergravity \eqref{NHnonExtD1} is now a solution of type IIA supergravity that describes a collection of D0-branes uniformly smeared along the transverse circle $\widetilde{S}^1$ with length $\widetilde L$ parametrised by $\widetilde{\theta}$.  The thermodynamic quantities of this solution are still given by \eqref{ThermoNHD1}.  Note that even after T-duality, the corresponding SYM theory still lies in the same manifold $\mathbb R^{(t)}\times S^1$. 
 
Consider now the validity of the type IIA supergravity description  \eqref{NHnonExtD0}. Smallness of the curvature in string units still requires that $\tau\ll \sqrt{\lambda^\prime}$. However, the requirement that the dilaton \eqref{NHnonExtD0}  at the horizon is small now yields $\tau\gg \lambda^{\prime\,5/18}/N^{4/9}$.  As for the requirements from the circle size, the mass of winding modes now goes as $M_w\sim \lambda^{1/4}/(\ell_s u_0^{3/2})$, so the condition $M_w/\sqrt{R}\gg 1$ and  \eqref{ThermoNHD1} now lead to the requirement that $\tau \ll 1$.  Additionally, avoiding string excitations from modes with momentum along the circle requires $\tau \ll \lambda^{\prime\,-1/6}$. 

Altogether, removing redundant bounds, we have
\begin{equation}\label{TdualD1valid}
\lambda^{\prime\,5/18}/N^{4/9}\ll \tau\;,\qquad\tau\ll \lambda^{\prime-1/6}\;,\qquad \tau\ll \sqrt{\lambda^\prime}\;.
\end{equation}
Note that as long as $N$ is sufficiently large, the IIA description is valid for much lower temperatures than the IIB description.

The D0-brane configuration \eqref{NHnonExtD0}  uniformly smeared along the  $\widetilde{S}^1$ circle exhibits a Gregory-Laflamme (GL) instability \cite{Gregory:1993vy,Gregory:1994bj,Susskind:1997dr,Barbon:1998cr,Li:1998jy,Aharony:2004ig,Harmark:2005jk} that will be the focus of this work. When $u_0\ll 1$, the D0-brane has a separation of length scales where the horizon radius is much smaller than the circle length $\widetilde{L}$. Such a configuration is unstable to deformations that break the symmetries of the $\widetilde{S}^1$. Note that this GL physics can be addressed only in the type IIA description since, from \eqref{ThermoNHD1}, $u_0\lesssim{\cal O}(1)$ implies that $\tau\lesssim {\cal O}(1/\sqrt{\lambda^\prime})$, which lies outside the validity of type IIB.  

\begin{figure}[th]
\centering
\includegraphics[width=.85\textwidth]{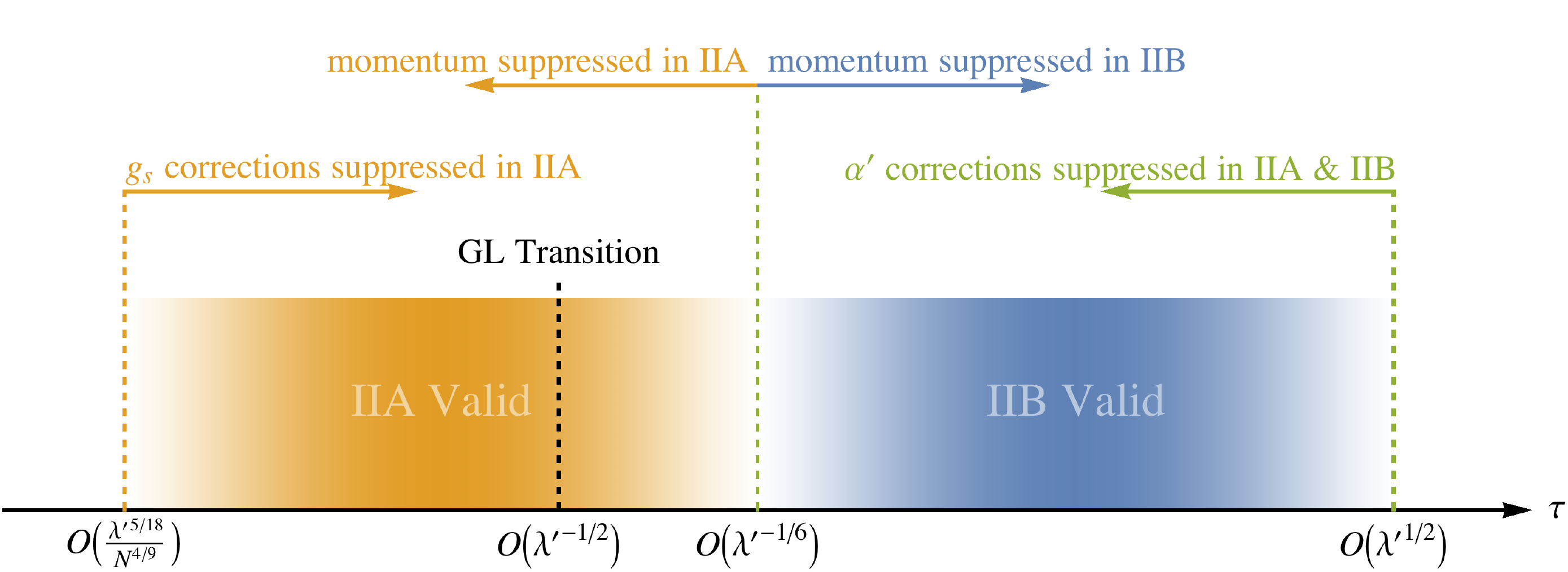}
\caption{Schematic regime of validity for IIA and IIB.  Here, we take $N^{4/7}\gg\lambda^\prime\gg1$, which is required for  the GL transition to be visible within supergravity.}\label{Fig:validitySUGRA}
\end{figure}

However, temperatures $\tau\lesssim {\cal O}(1/\sqrt{\lambda^\prime})$ are accessible within IIA for certain values of $\lambda^\prime$ and $N$.  Our bounds \eqref{TdualD1valid} require $N^{4/7}\gg\lambda^\prime\gg1$.  These criteria as well as the bounds of IIA and IIB are illustrated in Figure \ref{Fig:validitySUGRA}.

This instability was explicitly confirmed in \cite{Aharony:2004ig}, where the zero-mode onset of the instability was found to be at $u_0{\bigl |_{GL}}=(2\pi)^2 a_0$ or, equivalently, at\footnote{The value found in \cite{Aharony:2004ig} was $a_0\approx 0.37$. The value in \eqref{TGL} improves the accuracy.} 
\begin{equation}\label{TGL}
 \tau_{GL}=\frac{3}{4\sqrt{\pi}}\frac{(2\pi a_0)^2}{\sqrt{\lambda^{\prime}}}=\frac{2.243}{\sqrt{\lambda^{\prime}}}, \quad \hbox {for} \quad a_0\approx 0.3664.
\end{equation}
The presence of this zero mode implies the existence of new stationary solutions that break the symmetries of the $\widetilde{S}^1$ \cite{Aharony:2004ig}. Such solutions should describe a collection of D0-branes that are non-uniformly smeared along the circle.  From studies of similar systems with GL instabilities  \cite{Kol:2002xz,Kol:2003ja,Asnin:2006ip}, it is expected that if one continues along this new branch of solutions, the deformations along the $\widetilde{S}^1$ grow and the $S^7\times \widetilde{S}^1$ topology eventually changes to $S^8$.  After the topology change, the solution should describe D0-branes that are localised on the $\widetilde{S}^1$. There are thus three solutions that compete in a given thermodynamic ensemble: the uniform phase, the nonuniform phase, and the localised phase. Again, previous studies of GL instabilities \cite{Harmark:2002tr,Wiseman:2002ti,Kudoh:2003ki,Kudoh:2004hs,Sorkin:2006wp,Kleihaus:2006ee,Harmark:2007md,Headrick:2009pv,Wiseman:2011by,Figueras:2012xj,HorowitzBook2012,Kalisch:2016fkm} suggest that the localised phase is dominant at small energies and temperatures, while the uniform phase is dominant for larger energies and temperatures, and the phase transition is first order. However, without the nonuniform phase and the localised phases, these expectations remain heuristic and unverified.

In the (1+1)-dimensional gauge theory on the circle $S^1$, the GL instability is dual to spontaneously symmetry breaking. It has also been proposed that the gravitational first order phase transition between the uniform and localised phases on the circle $S^1$ is dual to localisation of the eigenvalue distribution associated to a Polyakov loop on the $S^1$ \cite{Aharony:2004ig,Catterall:2010fx}. If these expectations prove correct, the gravitational computation gives the critical value for the first order phase transition, at strong coupling, which can be tested with the critical value computed using SYM lattice computations (already started in \cite{Aharony:2004ig,Catterall:2010fx}). 

\section{Map between gravitational vacuum solutions and SYM thermal states \label{MapGRtoSYM}}

\subsection{Generating type IIA solutions from vacuum Einstein gravity solutions}
Rather than obtaining solutions to type IIA supergravity by directly solving the equations of motion, we will first solve the much simpler vacuum Einstein equation $R_{ab}=0$ in 10 dimensions, and then generate a solution to type IIA supergravity via an uplift-boost-KK reduction procedure which we will now review (see e.g. \cite{Aharony:2004ig,Harmark:2004ws}). 

We begin with any static, axially symmetric solution of vacuum Einstein gravity $R_{ab}=0$, that is asymptotically ${\mathbb R}^{(1,8)}\times \widetilde{S}^1$. Such a solution can be written in isotropic coordinates where the metric takes the form
\begin{equation}
{\mathrm d\bar{s}}^{\,2} = - A^2 {\mathrm d}T^2 + V^2 \left( {\mathrm d}\rho^2 + \rho^2
{\mathrm d}\Omega^2_{7} \right) + B^2 {\mathrm d}\widetilde{x}^{\,2},
\label{soln:vacuum10}
\end{equation}
where $\widetilde{x}\sim \widetilde{x}+{\widetilde L}$ parametrises the compact circle $\widetilde{S}^1$ with length $\widetilde{L}$, and $A, V, B$ are functions of $\rho, \widetilde{x}$ which approach unity at large radial direction $\rho$. We further assume that the solution is a black hole and therefore $A$ vanishes at the horizon location, $\rho=\rho_0$.

This solution can be uplifted to 11-dimensions with an extra $Z$ coordinate via ${\mathrm ds}^2_{11}={\mathrm d\bar{s}}^2+{\mathrm d}Z^2$. We can now boost this solution along the $Z$-direction, $T=\cosh\beta\, t+\sinh\beta\, z, Z=\sinh\beta\, t+\cosh\beta\, z$, which gives  
\begin{equation}
{\mathrm ds}^2_{11}= e^{-\frac{1}{6}(\tilde{\phi}-\tilde{\phi}_\infty)}{\mathrm ds}^2_{(E)}+e^{\frac{4}{3}(\tilde{\phi}-\tilde{\phi}_\infty)} \left( {\mathrm d}z-\widetilde{A}_t {\mathrm d}t \right)^2, 
\label{soln:11boost}
\end{equation}
with ${\mathrm ds}^2_{(E)}= e^{-\frac{1}{2}(\tilde{\phi}-\tilde{\phi}_\infty)} {\mathrm d\widetilde{s}}^{\,2}$, and $\tilde{\phi}$ and $\widetilde{A}_t$ are to be defined in \eqref{soln:10KK} below. This procedure yields a solution of vacuum Einstein gravity in 11 dimensions.

Finally, we can do a Kaluza-Klein reduction along the $z$-direction. This yields a solution of type IIA supergravity action \eqref{action} with $p=0$, where the graviton, dilation, and 1-form field are given (in the string frame) by\footnote{One could also T-dualise \eqref{soln:10KK} to get a solution of type IIA supergravity action \eqref{action} with $p=1$.}
\begin{eqnarray}
&& {\mathrm d \widetilde{s}}^{\,2}= \left(\cosh^2\beta - A^2 \sinh^2\beta \right)^{1/2} \left[ - \frac{A^2}{\cosh^2\beta - A^2 \sinh^2\beta} {\mathrm d}t^2 + V^2 \left( {\mathrm d}\rho^2 + \rho^2
{\mathrm d}\Omega^2_{7} \right) + B^2 {\mathrm d}\widetilde{x}^2\right],\nonumber \\
&& e^{\widetilde{\phi}}= \widetilde g_s \left(\cosh^2\beta - A^2 \sinh^2\beta \right)^{3/4}\,,\qquad  \widetilde{A}_t=-\frac{1}{\widetilde g_s}\frac{\left(A^2-1\right)\cosh\beta \sinh\beta}{\cosh^2\beta - A^2 \sinh^2\beta}\,.
\label{soln:10KK}
\end{eqnarray}

The original vacuum solution \eqref{soln:vacuum10} of 10-dimensional vacuum Einstein gravity can be a static uniform black string, a nonuniform black string or a localised black hole on the circle $\widetilde{S}^1$. In each of these three cases, the uplift-boost-KK reduction procedure above generates an associated uniform, nonuniform or localised solution of type IIA supergravity with a dilaton and 1-form.

As an example, let us obtain the type IIA uniform distribution of D0-branes \eqref{NHnonExtD0} through this procedure. The line element for the asymptotically ${\mathbb R}^{(1,8)}\times \widetilde{S}^1$ uniform black string is just the product of a circle with a Schwarzschild-Tangherlini black hole: 
\begin{equation}\label{uniformstring}
{\mathrm d\bar{s}}^{\,2} = - f {\mathrm d}T^2 + \frac{{\mathrm d}r^2}{f} + r^2 {\mathrm d}\Omega^2_{7}  + {\mathrm d}\widetilde{x}^{\,2}\;,\qquad f=1-\frac{r_0^6}{r^6}\;.
\end{equation}
Under the coordinate transformation $r=\rho \left( 1+\frac{\rho_0^6}{\rho^6}\right)^{1/3}$ and $r_0=2^{1/3} \rho_0$, this metric can be brought to isotropic form \eqref{soln:vacuum10} with
\begin{equation} \label{soln:10startUnif}
A=\frac{\rho^6-\rho_0^6}{\rho^6 + \rho_0^6}\,, \quad B=1\,, \quad V=\left( 1+\frac{\rho_0^6}{\rho^6}\right)^{1/3}.
\end{equation}
Applying the uplift-boost-KK reduction procedure, we find that \eqref{soln:10KK} reduces in this particular case to
\begin{eqnarray}
&& {\mathrm d\widetilde{s}}^{\,2}=  - H^{-\frac{1}{2}}f {\mathrm d}t^2 + H^{\frac{1}{2}} \left( \frac{{\mathrm d}r^2}{f} + r^2
{\mathrm d}\Omega^2_{7}  + {\mathrm d}\widetilde{x}^{\,2}\right),\qquad e^{\widetilde{\phi}}=\widetilde g_s\,H^{3/4}\,,\qquad  \widetilde{A}_{t}=-{\widetilde g_s}^{-1} \hbox{coth}\beta \left(H^{-1} -1 \right)\,,\nonumber \\
&& f=1-\frac{r_0^6}{r^6},\qquad H=1+\frac{r_0^6}{r^6}\sinh^2\beta \,,
\label{soln:10KKUnif}
\end{eqnarray}
This IIA solution is the full geometry of the near-horizon solution \eqref{NHnonExtD0} that describes a collection of D0-branes uniformly smeared along the transverse circle $\widetilde{S}^1$ parametrised by $\widetilde{x}$.
To check this is indeed the case apply to \eqref{soln:10KKUnif} the transformations  $r=\frac{\ell_s^2}{L}\,u$, $r_0=\frac{\ell_s^2}{L}\,u_0$, $\widetilde{x}=\frac{\ell_s^2}{L}\,2\pi \widetilde{\theta}$ and the identification $\sinh \beta = \sqrt{d_1 \lambda^{\prime}}L^2/(\ell_s^2 u_0^3)$ which leads to
\begin{eqnarray}\label{FullnonExtD0}
&& {\mathrm d\widetilde{s}}^{\,2}= \ell_s^2\left[-\left( \frac{\ell_s^4}{L^4} +\frac{d_1 \lambda^{\prime}}{u^6}\right)^{-\frac{1}{2}} \left( 1-\frac{u_0^6}{u^6} \right) \frac{{\mathrm d}t^2}{L^2} +\left( \frac{\ell_s^4}{L^4} +\frac{d_1 \lambda^{\prime}}{u^6}\right)^{\frac{1}{2}} \left(\left( 1-\frac{u_0^6}{u^6} \right)^{-1} {\mathrm d}u^2 +u^2 {\mathrm d}\Omega_{(7)}^2 + (2 \pi )^2{\mathrm d}\widetilde{\theta}^{\,2} \right) \right],  \nonumber \\
&&  e^{\widetilde{\phi}}=(2 \pi)^2 \frac{\lambda^{\prime} }{N} \left( \frac{\ell_s^4}{L^4} +\frac{d_1 \lambda^{\prime}}{u^6}\right)^{\frac{3}{4}}\!,
   \qquad   \widetilde{A}_{(1)}=- \frac{1}{(2\pi)^2} \frac{L^3}{\ell_s^3} \frac{u_0^3}{\sqrt{d_1} \sqrt{\lambda^\prime}} \frac{N}{\lambda^\prime} \left( \frac{\ell_s^4}{L^4} +\frac{d_1 \lambda^{\prime}}{u_0^6}\right)^{\frac{1}{2}} \left[ \frac{\ell_s^4}{L^4}  \left( \frac{\ell_s^4}{L^4} +\frac{d_1 \lambda^{\prime}}{u^6}\right)^{-1}-1\right]\,{\mathrm d}t,  \nonumber \\
&& 
\end{eqnarray}  
where we have used the T-dual relation \eqref{Tduality} as well as the definitions \eqref{gYM} and \eqref{lambdaprimedef} to replace $\widetilde g_s$. The identification $\sinh \beta = \sqrt{d_1 \lambda^{\prime}}L^2/(\ell_s^2 u_0^3)$ can be understood in the following manner: by applying a T-duality to (\ref{soln:10KKUnif}) we get the metric corresponding to a stack of $N$ coincident $D1$ branes in flat space; we then use (\ref{defKp}) with $p=1$. Finally, from \eqref{FullnonExtD0}, gauging away a constant term in $\widetilde{A}_{(1)}$ and taking the decoupling limit $\ell_s\to 0$ yields \eqref{NHnonExtD0}.

For the non-uniform and localised solutions on the circle $\widetilde{S}^1$, the vacuum Einstein field equations can only be solved numerically or, in the localised case for small energies, within perturbation theory \cite{Harmark:2003yz,Dias:2007hg}. Nevertheless, once these vacuum solutions are constructed, the uplift-boost-KK reduction procedure can be used to  generate the associated IIA partner.

It is worth emphasising that the uniform solution \eqref{FullnonExtD0}, like others that we will generate in this way, is asymptotically ${\mathbb R}^{(1,8)}\times \widetilde{S}^1$ and therefore corresponds to the full geometry rather than the near-horizon limit \eqref{NHnonExtD0}. The near-horizon limit is dual to the decoupling limit that introduces the SYM theory. Therefore, in order to properly compare supergravity results with SYM, we must take the near-horizon decoupling limit \eqref{NHlim}. We further note that the zero mode of the Gregory-Laflamme instability \eqref{TGL} was also obtained \cite{Aharony:2004ig} studying perturbations of the full geometry \eqref{FullnonExtD0} rather than from the near-horizon limit \eqref{NHnonExtD0}. This is justified by the fact that the zero mode in the full geometry has support mostly near the horizon and decays exponentially. So, it is therefore likely to lie within the near-horizon geometry. 

\subsection{SYM thermodynamics from the thermodynamics of vacuum solutions}
Now we explain how to obtain the thermodynamic quantities of thermal states of SYM on $\mathbb R^{(t)}\times S^1$ from a given asymptotically  ${\mathbb R}^{(1,8)}\times \widetilde{S}^1$ solution \eqref{soln:vacuum10} to the 10-dimensional vacuum Einstein equation \cite{Aharony:2004ig,Harmark:2004ws}. The process, briefly, begins with performing the uplift-boost-KK reduction procedure to a vacuum solution in the form \eqref{soln:vacuum10} to obtain the thermodynamic quantities of the corresponding type IIA supergravity solution on the circle $\widetilde{S}^1$.  We then T-dualise to get the thermodynamics of the associated IIB supergravity solution on the ${S}^1$. Finally, we take the decoupling limit \eqref{NHlim} which yields the thermodynamics of the corresponding near-horizon solution. From there, we can obtain SYM thermodynamic quantities through the duality. 

Applying the uplift-boost-KK reduction procedure to the vacuum solution \eqref{soln:vacuum10} one gets a static type IIA solution \eqref{soln:10KK} with a RR 1-form that asymptotes to ${\mathbb R}^{(1,8)}\times \widetilde{S}^1$. 
A given family of such solutions can be parametrised by a dimensionless quantity $\xi_0=\rho_0/\widetilde{L}$, where $\rho=\rho_0$ is the location of the horizon and  $\widetilde{L}$ is the length of the circle $\widetilde{S}^1$, and by the RR gauge (KK boost) parameter $\beta$. The ${\mathbb R}^{(1,8)}\times \widetilde{S}^1$ asymptotics imply that the functions $A(\rho,\widetilde{x})$ and $B(\rho,\widetilde{x})$ in \eqref{soln:vacuum10}  or \eqref{soln:10KK} admit an asymptotic Taylor expansion around $\rho=\infty$ of the form,
\begin{equation}\label{10KKdecay}
A=  1 - a(\xi_0)  \frac{\widetilde{L}^{6}}{\rho^{6}}  + \cdots\,, \qquad\qquad
B= 1 + b(\xi_0) \frac{\widetilde{L}^{6}}{\rho^{6}}  + \cdots\;,
\end{equation}
where (for a given $\xi_0$) $a, b$ are integration constants that depend on the solution at hand, \emph{i.e.} they cannot be determined by an expansion at infinity alone and are fixed after solving the equations of motion subject to some interior boundary condition (in the present case, regularity at the horizon).  The decay of the remaining function $V$ is fixed by $a$ and $b$. 

The conserved ADM charges, namely the mass $M$ and the electric charge $Q$,  
of the static solution \eqref{soln:10KK} are then a function of $a, b$ and $\beta$ \cite{Kol:2003if,Harmark:2003dg,Harmark:2003yz,Aharony:2004ig,Harmark:2004ws},\footnote{In \eqref{10KKadm} and \eqref{10KKhorizon} the extensive quantities $M/\widetilde{L}$, $Q/\widetilde{L}$ and $S/\widetilde{L}$ give, respectively, the mass, charge and entropy densities along the transverse circle $\widetilde{S}^1$. The extra powers of $\widetilde{L}$ follow from working with the adimensional horizon radius $\xi_0=\rho_0/\widetilde{L}$. Further note that setting $\beta=0$ in  \eqref{10KKadm} and \eqref{10KKhorizon} we get the thermodynamics of the vacuum Einsteins solution \eqref{soln:vacuum10}.}
\begin{eqnarray}
\label{10KKadm}
M = \frac{\widetilde{L}^{7}\, \Omega_{7}}{8 \pi G_{10}} {\biggl (} 7 a(\xi_0) -  b(\xi_0) +  6 \,a(\xi_0) \sinh^2 \beta {\biggr )} , \qquad 
Q = \frac{\widetilde{L}^{7}\, \Omega_{7}}{8 \pi G_{10}} \, 3 \,a(\xi_0) \sinh 2 \beta .
\end{eqnarray}
The entropy, temperature and chemical potential can be obtained from $\beta$ and horizon quantities \cite{Kol:2003if,Harmark:2003dg,Harmark:2003yz,Aharony:2004ig,Harmark:2004ws}
\begin{equation}\label{10KKhorizon}
S = \frac{\widetilde{L}^8\, \Omega_{7}}{4 G_{10}} \cosh\beta \, s_h(\xi_0), \qquad
T =  \frac{1}{\widetilde{L}}  \frac{ t_h(\xi_0)}{\cosh\beta}, \qquad \mu = \tanh\beta \,, 
\end{equation}
where
\begin{equation}\label{stdef}
s_h=\xi_0^7 B(\xi_0) V(\xi_0)^7, \qquad t_h=\frac{A'(\xi_0)}{2 \pi  V(\xi_0)}.
\end{equation}
Fixing the length of the asymptotic circle $\widetilde{L}$ and $\beta=0$, we  recover the thermodynamic quantities of the 10-dimensional vacuum Einstein solution where quantities obey the first law and Smarr relations \cite{Kol:2003if,Harmark:2003dg,Harmark:2003yz,Aharony:2004ig,Harmark:2004ws}:
\begin{equation}\label{10GR1sttlawSmarr}
\mathrm dM(\xi_0,0) = T(\xi_0,0) \,\mathrm dS(\xi_0,0)\,, \qquad\qquad \frac{\widetilde{L}^7\,\Omega_{7} }{4 \pi G_{10}}\, 3 \,a(\xi_0)  =  T(\xi_0,0) \,S(\xi_0,0).
\end{equation}
 It follows that \eqref{10KKadm} and \eqref{10KKhorizon}  obey the first law of thermodynamics
\begin{equation}\label{10KKfirstlaw}
\mathrm dM(\xi_0,\beta) = T(\xi_0,\beta) \, \mathrm dS(\xi_0,\beta) + \mu(\beta) \,dQ(\xi_0,\beta).
\end{equation}

At this stage we have the thermodynamic quantities $\{M,Q,S,T,\mu\}$ in the full geometry \eqref{soln:10KK} of IIA supergravity. We now apply the T-duality \eqref{Tduality} and take the decoupling limit \eqref{NHlim} to obtain the thermodynamics $\{\varepsilon,\sigma,\tau\}$ of the associated IIB near-horizon geometry. 
Moreover, according to the holographic conjecture of \cite{Itzhaki:1998dd,Aharony:2004ig}, $\{\varepsilon,\sigma,\mathcal{T}\}$ describe also the thermodynamics  of thermal states on the dual SYM theory on $\mathbb R^{(t)}\times S^1$. 

More concretely, to obtain the above dictionary $\{M,Q,S,T,\mu\} \to \{ \varepsilon, \sigma, \tau \}$ we apply the decoupling limit \eqref{NHlim}\footnote{Note that the charge computed in (\ref{10KKadm}) is that of a distribution of D$0$, while we need to keep fixed in the decoupling limit the $D1$ brane charge.} and the T-duality relations \eqref{Tduality} for the circle length $\widetilde{L}(L)$ and string coupling $\widetilde{g}_s(g_s)$. In this process, recall that \eqref{10KKadm}-\eqref{10KKhorizon} correspond to type IIA quantities, where Newton's constant translates to field theory language as $16 \pi  G_{10}\equiv (2 \pi )^7 \widetilde{g}_s^{\,2} \ell_s^8$. Also recall that the string coupling of IIB with a RR 2-form is given in terms of the SYM coupling by $g_s\equiv 2 \pi \ell_s^{2} g_{\rm YM}^2$.  Altogether, we find that the SYM dimensionless energy, entropy and temperature are: 
\begin{equation}
\label{ThermoSYM}
\varepsilon=\frac{16 \pi^7}{3} \, {\bigl [}4 a(\xi_0)-b(\xi_0) {\bigr ]} \,  \frac{N^2}{ \lambda^{\prime\,2}}, \qquad  \qquad  \sigma=\frac{16 \pi^{11/2}}{3} \,\frac{s_h(\xi_0)}{\sqrt{2 a(\xi_0)}} \,\frac{N^2}{ \lambda^{\prime\,\frac{3}{2}}}, \qquad  \qquad  \tau=2 \pi^{5/2}\sqrt{2 a(\xi_0)}\, t_h(\xi_0)\,\frac{1}{ \sqrt{\lambda^{\prime}} }.
\end{equation}
Given a vacuum Einstein gravity solution in isotropic coordinates \eqref{soln:vacuum10} that asymptotes to ${\mathbb R}^{(1,8)}\times \widetilde{S}^1$, we can read the parameters $a(\xi_0)$ and $b(\xi_0)$ from the asymptotic decay \eqref{10KKdecay} of the functions $A(\rho,\widetilde{x})$ and $B(\rho,\widetilde{x})$ and the parameters $s_h(\xi_0)$ and $t_h(\xi_0)$ from the horizon via \eqref{10KKhorizon} and \eqref{stdef}. However, it is not always practical to obtain a solution in isotropic coordinates. It is thus desirable to write the SYM quantities \eqref{ThermoSYM} in terms of gauge invariant gravitational quantities. First, we obtain the mass $M$, entropy $S$, and temperature $T$ within 10-dimensional vacuum Einstein using standard ADM techniques.  We can then match these with the expressions \eqref{10KKadm} and \eqref{10KKhorizon} (with $\beta=0$). Together with the Smarr law \eqref{10GR1sttlawSmarr}, this matching gives four equations  to solve for $a(\xi_0),b(\xi_0),s_h(\xi_0)$ and $t_h(\xi_0)$. Inserting these into \eqref{ThermoSYM} gives the SYM thermodynamical quantities
\begin{equation}\label{map}
\varepsilon=64\pi^4 \left(2\widehat M-\widehat S\, \widehat T \right)\, \frac{N^2}{{\lambda^\prime}^2}\;,\qquad \sigma=16\sqrt{2}\pi^3\sqrt{\frac{\widehat S}{\widehat T}}\,
\frac{N^2}{{\lambda^\prime}^{3/2}}\;,\qquad \tau=4\sqrt 2 \,\pi \,\widehat S^{\,1/2} \,\widehat T^{\,3/2}\,\frac{1}{\sqrt{\lambda^\prime}}\;,
\end{equation}
where
\begin{equation}\label{dimlessthermo}
\widehat M=\frac{G_{10} M}{ \widetilde{L}^7}\;,\qquad \widehat S=\frac{G_{10} S}{ \widetilde{L}^8}\;,\qquad \widehat T= \widetilde{L}\,T
\end{equation}
are the dimensionless gravitational mass, entropy, and temperature, respectively. 

To summarise, given the thermodynamics of a vacuum asymptotically ${\mathbb R}^{(1,8)}\times \widetilde{S}^1$ gravitational solution, the thermodynamics of the dual SYM theory follows directly from the map \eqref{map}.  We can therefore bypass the type IIA equations of motion entirely and solve the simpler vacuum Einstein equations. Once the uniform, nonuniform, and localised phases are available, the preferred phase in a given thermodynamic ensemble can be determined by comparing thermodynamic potentials through the map \eqref{map}.

\subsection{Thermodynamics for the uniform and perturbative localised phases}
\label{ThermoUnifPert}

Let us apply the map discussed in the previous two subsections to phases for which one has an analytical or perturbative solution. 
The only asymptotically ${\mathbb R}^{(1,8)}\times \widetilde{S}^1$ vacuum Einstein solution known entirely in closed form is the uniform black string, which has horizon topology horizon topology $S^7\times \widetilde{S}^1$.  Its line element in isotropic coordinates \eqref{soln:vacuum10} was given by \eqref{soln:10startUnif}. 
 Then, from a series expansion at infinity \eqref{10KKdecay}, we find that $a=2\xi_0^6$ and $b=0$. At the horizon, we find from the definitions \eqref{stdef} that $s_h=2^{7/3} \xi_0^7$ and $t_h=3/(2^{4/3}\pi \xi_0)$. 
Plugging this directly into  \eqref{10KKadm} and \eqref{10KKhorizon} one finds that the thermodynamics of the uniform black string branch is
\begin{eqnarray}
\label{ThermoGRunif}
&& \qquad M(\xi_0)=\frac{\widetilde{L}^{7}}{G_{10}}\,\frac{7 \,\pi^3}{12}\,\xi_0^6 , \qquad \qquad
S(\xi_0)=\frac{\widetilde{L}^{8}}{G_{10}}\,\frac{2^{1/3} \pi^4}{3} \,\xi_0^7\,,\qquad \nonumber\\
&& \qquad  T(\xi_0)=\frac{1}{\widetilde{L}}\,\frac{3}{2^{4/3} \pi }\,\frac{1}{\xi_0}\,,\qquad  \qquad
F(\xi_0)=\frac{\widetilde{L}^{7}}{G_{10}}\,\frac{\pi^3}{12}\,\xi_0^6\,,
\end{eqnarray}
where $F=M-TS$ is the Helmoltz free energy.  For the purpose of later presenting our results in Section \ref{sec:ResultsGR}, it will be useful for us to express the entropy as a function of the energy and the free energy in terms of the temperature:
\begin{eqnarray}
S(M)=\frac{3^{1/6} 2^{8/3} \sqrt{\pi }}{7^{7/6}}\left(\frac{G_{10}}{\widetilde{L}^7}\,M \right)^{7/6}\frac{\widetilde{L}^{8}}{G_{10}},\qquad\qquad F(T)=\frac{243}{1024 \,\pi^3} \frac{1}{(T \widetilde{L})^6}\frac{\widetilde{L}^{7}}{G_{10}}\,.
\end{eqnarray}

However, we are mainly interested in the SYM thermodynamics dual to \eqref{ThermoGRunif}.  These have already been presented in \eqref{ThermoNHD1} which can be obtained directly from the uniform phase within type IIA or type IIB supergravity. As a check, following the procedure outlined in the previous subsection, i.e.  plugging the values of $\{a,b,s_h,t_h\}$ for the uniform solution into \eqref{ThermoSYM} gives exactly \eqref{ThermoNHD1}, as expected (this procedure is equivalent to applying the map \eqref{map} directly)\footnote{Note that $\xi_0=\frac{u_0}{2^{1/3}(2\pi)^2}$. This follows from $\xi_0=\rho_0/\widetilde{L}$ with $\rho_0=2^{-1/3}r_0$ and $r_0=u_0 \ell_s^2/L$.}.  Later, in Section \ref{sec:Results}, it will be useful to have the entropy as a function of the energy and the free energy $\mathfrak{f}=\varepsilon-\tau\sigma$ as a function of the temperature $\tau$ for this phase:
\begin{equation}
\label{ThermoSYMunif}
 \sigma(\varepsilon)=\frac{2^{2/3} \pi^{5/6}}{3^{1/3}}\,\left(\frac{\lambda^{\prime\,2}}{N^{2}}\,\varepsilon\right)^{2/3}\, \frac{N^{2}}{\lambda^{\prime\,3/2}}, \quad\qquad 
\mathfrak{f}(\tau)=-\frac{16}{81}\,\pi^{5/2}\, \left(\sqrt{\lambda^{\prime}}\,\tau\right)^3 \, \frac{N^2}{\lambda^{\prime\,2}}\,.
\end{equation}

Another asymptotically ${\mathbb R}^{(1,8)}\times \widetilde{S}^1$ solution of vacuum Einstein gravity is a black hole localised on the circle $\widetilde{S}^1$ with horizon topology $S^8$. When energies are low compared to the circle size, the geometry near the horizon resembles that of an asymptotically flat 10-dimensional Schwarzschild-Tangherlini black hole.  At larger energies, the presence of the circle deforms the horizon.  When these deformations are small, they can be captured perturbatively through an expansion in $ \xi_0=R_0/\widetilde{L} \ll 1$, where $R_0$ is the horizon radius and $\widetilde{L}$ is the size of $\widetilde{S}^1$ \cite{Harmark:2003yz,Gorbonos:2004uc,Harmark:2004ws,Dias:2007hg}. In particular, thermodynamic quantities (within vacuum Einstein) can be found in section 6 of \cite{Harmark:2003yz}:\footnote{Note that \cite{Harmark:2003yz} sets ${L}=2\pi$.  Here, we express their results in terms of the dimensionless parameter $\xi_0=R_0/{L}$. $\zeta (s)=\sum _{k=1}^{\infty } k^{-s}$ is the Riemann zeta function and $\Omega_8=\frac{32}{105}\,\pi ^4$.}
\begin{eqnarray}
\label{thermoLocGRpert}
&&  \qquad \qquad  M(\xi_0)= \frac{\widetilde{L}^7}{G_{10}} \frac{\Omega_{8}}{2\pi} \,\xi_0^{7} 
\left( 1 + \frac{\zeta(7)}{2} \,\xi_0^{7} 
+ \mathcal{O} ( \xi_0^{14} ) \right) \,,
\qquad  S(\xi_0)=  \frac{\widetilde{L}^8}{G_{10}} \frac{\Omega_{8}}{4}\, \xi_0^{8} 
\left( 1 + \frac{8\,\zeta(7)}{7}\, \xi_0^{7} 
+ \mathcal{O}  ( \xi_0^{14} ) \right),
\nonumber \\
&&  \qquad \qquad   T(\xi_0) = \frac{1}{\widetilde{L}}\frac{7}{4\pi\, \xi_0} 
\left( 1 - \frac{8\,\zeta(7)}{7} \xi_0^{7} 
+ \mathcal{O} ( \xi_0^{14} ) \right)\,,
\qquad F(\xi_0)= \frac{\widetilde{L}^7}{G_{10}} \frac{\,\Omega_{8}}{16\pi} \,\xi_0^{7} 
{\biggl (} 1 + 4\,\zeta(7) \,\xi_0^{7} 
+ \mathcal{O} ( \xi_0^{14} ) {\biggr )}  \,;\nonumber \\
 &&S(M)=\left(\frac{105 \pi ^4}{2^{11}}\right)^{1/7} \left(\frac{G_{10}}{\widetilde{L}^7}M\right)^{8/7}\left( 1+\frac{15 \zeta(7)}{4\pi^3}\frac{G_{10}}{\widetilde{L}^7}M +\mathcal{O}(M^2) \right)  \frac{\widetilde{L}^8}{G_{10}}, \nonumber\\
&& F(T)=\frac{117649}{122880\, \pi^4}\frac{1}{\widetilde{L}^7 T^7}\left( 1-\frac{823543 \,\zeta (7)}{4096 \,\pi^7} \frac{1}{(\widetilde{L} T)^7} +\mathcal{O}(T^{-14}) \right)\frac{\widetilde{L}^7}{G_{10}}\,.
\end{eqnarray}
The dual SYM quantities can be obtained through the map \eqref{map}:
\begin{eqnarray}
\label{ThermoSYMpertLoc}
&& \qquad \qquad \varepsilon(\xi_0)= \frac{384 \pi ^7}{35}\,\xi_0^7\left(1+ \frac{8 \zeta (7)}{9} \xi_0^7 + \mathcal{O} ( \xi_0^{14} ) \right) \frac{N^2}{ \lambda^{\prime\,2}}\,,
\qquad  \sigma(\xi_0)=  \frac{128 \pi ^{11/2}}{7 \sqrt{15}}\,\xi_0^{9/2}\left(1+
\frac{8 \zeta (7)}{7}\,\xi_0^7+ \mathcal{O}  ( \xi_0^{14} ) \right)\frac{N^2}{ \lambda^{\prime\,\frac{3}{2}}},
\nonumber \\
&& \qquad  \qquad  \tau(\xi_0) =\frac{14 \pi ^{3/2}}{\sqrt{15}} \,\xi_0^{5/2}\left(1-\frac{8 \zeta (7)}{7}\,
\xi_0^7+ \mathcal{O} ( \xi_0^{14} )\right)\frac{1}{ \sqrt{\lambda^{\prime}} }\,,
\qquad \mathfrak{f}(\xi_0)= -\frac{128 \pi ^7}{21} \,\xi_0^7 \left(1-\frac{8 \zeta (7)}{5} \xi_0^7+ \mathcal{O} ( \xi_0^{14} )\right)  \frac{N^2}{ \lambda^{\prime\,2}}  \,;\nonumber \\
 &&\sigma(\varepsilon) =\frac{2^{5/2} 5^{1/7} \pi }{3^{8/7} 7^{5/14}}\,\left(\frac{ \lambda^{\prime\,2}}{N^2} \,\varepsilon\right) ^{9/14}\left( 1+  \frac{5 \,\zeta (7)}{96 \pi ^7}\frac{ \lambda^{\prime\,2}}{N^2}\, \varepsilon  +\mathcal{O}(\varepsilon^2) \right)\frac{N^2}{ \lambda^{\prime\,\frac{3}{2}}}, \nonumber\\
&& \mathfrak{f}(\tau)=-\frac{2^{1/5} 15^{2/5} 80\,\pi^{14/5}}{7^{4/5}343}\,\left(\sqrt{\lambda^{\prime}} \, \tau\right)^{14/5}\left[ 1+ \frac{2^{1/5} 15^{2/5} 3\,\zeta (7)}{7^{4/5} 49\,\pi^{21/5}} \left(\sqrt{\lambda^{\prime}} \, \tau\right)^{14/5} +\mathcal{O}(\tau^{28/5})\right]  \frac{N^2}{ \lambda^{\prime\,2}}\,.
\end{eqnarray}
For larger energies, the localised solutions can only be obtained numerically, which we will be done in the next section. In Section \ref{sec:Results} we will compare our numerical results with the perturbative results \eqref{ThermoSYMpertLoc}. We will find that the expressions \eqref{ThermoSYMpertLoc} give an excellent approximation, even beyond the regime where they may be expected to be valid.

\section{Non-uniform and Localised Phases} \label{sec:NumConstruction}
Now we numerically construct the remaining non-uniform and localised phases.  For this section, we will stay within the language of 10-dimensional vacuum Einstein gravity with ${\mathbb R}^{(1,8)}\times \widetilde{S}^1$ asymptotics where the $\widetilde{S}^1$ has circumference $\widetilde{L}$. Afterwards, in  Section \ref{sec:Results}, we use the results of Section \ref{MapGRtoSYM} and its map \eqref{map} to read off the thermodynamics of the dual SYM theory on $\mathbb R^{(t)}\times S^1$. 

Our numerical formalism of choice is the DeTurck method \cite{Headrick:2009pv,Figueras:2011va,Wiseman:2011by,Dias:2015nua}. This method requires that we first choose a reference metric $\overline g$.  This metric need not be a solution to the Einstein equation, but must contain the same symmetries and causal structure as the desired solution.  With the reference metric chosen, the DeTurck method then modifies the Einstein equation $R_{\mu\nu}=0$ to
\begin{equation}\label{EdeT}
R_{\mu\nu}-\nabla_{(\mu}\xi_{\nu)}=0\;,\qquad \xi^\mu \equiv g^{\alpha\beta}[\Gamma^\mu_{\alpha\beta}-\overline{\Gamma}^\mu_{\alpha\beta}]\;,
\end{equation}
where $\Gamma$ and $\overline{\Gamma}$ define the Levi-Civita connections for $g$ and $\bar g$, respectively. Unlike $R_{\mu\nu}=0$, this equation yields PDEs that are elliptic in character. But after solving these PDEs, we must verify that $\xi^\mu=0$ to confirm that $R_{\mu\nu}=0$ is indeed solved.\footnote{The condition $\xi^\mu=0$ also fixes all gauge freedom in the metric.}  Fortunately, the results of \cite{Figueras:2011va} have proven that static solutions to \eqref{EdeT} must satisfy $\xi^\mu=0$. Nevertheless, we will still monitor $\xi^\mu$ as a measure of numerical accuracy. 

\subsection{Nonuniform black strings} \label{sec:SetupNonUnif}
The non-uniform black strings we seek are asymptotically ${\mathbb R}^{(1,8)}\times \widetilde{S}^1$ and have horizon topology $S^7\times{S}^1$.  They are static and axisymmetric, and so only depend upon a periodic coordinate $\chi$ and a radial coordinate $y$. Since the uniform black string solution has the same symmetries and causal structure, we are free use it as a reference metric to find the non-uniform strings. We choose the reference metric
\begin{equation}
{\overline{\mathrm ds}}^2 = {\widetilde L}^2 {\biggl (} -G \, y^2\, {\mathrm d}t^2 +
\frac{4 \, y_+^2 \,{\mathrm d}y^2}{G \left(1-y^2\right)^4}      
 +{\mathrm d}\chi^2   + \frac{y_+^2}{\left(1-y^2\right)^2} \,{\mathrm d}\Omega_7^2
{\biggr )}\;,
\end{equation}
where
\begin{equation}\label{NonUnif:ansatz2}
     G=\left(2-y^2\right) \left(y^4-3 y^2+3\right) \left(y^4-y^2+1\right).
\end{equation}
We can arrive at this line element by taking the the black string the more familiar coordinates of \eqref{uniformstring}, and performing the redefinitions $r=y_+/(1-y^2)$, $T=\widetilde Lt$, $x=\widetilde L\chi$, $r_0=y_+$.  In these coordinates, $y\in[0,1]$, and $\chi\in[0,1)$ is periodic.  

We can now use this reference metric to create a metric ansatz
\begin{equation}\label{NonUnif:ansatz}
      {\mathrm ds}^2 = {\widetilde L}^2 {\biggl (} -G \, y^2 \,  q_1 \, {\mathrm d}t^2 +
\frac{4\, q_2 \, y_+^2 \,{\mathrm d}y^2}{G \left(1-y^2\right)^4}      
 +q_4 ( {\mathrm d}\chi+ q_3 \, {\mathrm d}y)^2   + \frac{y_+^2}{\left(1-y^2\right)^2} \,q_5 \,{\mathrm d}\Omega_7^2 
{\biggr )}\;,
\end{equation}
where, $q_i$ are unknown functions of $\chi$ and $y$. This ansatz is the most general form allowed by the symmetries.  With a reference metric and ansatz, we can now solve the Einstein-DeTurck equation \eqref{EdeT} subject to the following boundary conditions.  Horizon regularity is required at $y=0$ where we impose $q_1{\bigl |}_{y=0}=q_2{\bigl |}_{y=0}$ and Neumann conditions $\partial _yq_i{\bigl |}_{y=0}=0$ for $i\neq1$. ${\mathbb R}^{(1,8)}\times \widetilde{S}^1$ asymptotics requires that the metric be the same as the reference metric at $y=1$. Lastly, while $\chi$ is periodic in $\chi\in[0,1)$, the expected non-uniform string also has a $\mathbb Z_2$ symmetry in this coordinate.  We make use of this symmetry by taking $\chi\in[0,1/2]$ and imposing Neumann conditions for all $q_i$ at $\chi=0$ and $\chi=1/2$. 

We solve these equations numerically using a Newton-Raphson algorithm.  Discretisation is done using pseudospectral methods with Chebyshev-Gauss-Lobatto grids, and the resulting linear equations are solved using LU decomposition. To get a first seed, we added the GL linear perturbation to the uniform black string solution \eqref{uniformstring}.

The entropy and temperature of this black string are
\begin{equation}\label{thermoNonUnifGRnum}
S=\frac{{\widetilde L}^8}{G_{10}} \,\frac{\pi ^4}{6}\,y_+^7\,\int_0^{1/2}\mathrm d\chi\,\sqrt{q_4(\chi,0)} \,q_5(\chi,0)^{7/2},\qquad\qquad T=\frac{1}{{\widetilde L}}\,\frac{3}{2 \pi } \frac{1}{y_+}\;.
\end{equation}
and we see that $y_+$ directly determines the temperature and parametrises our family of solutions. (Our equations of motion do not depend on $\widetilde L$ which just sets a scale.)  To obtain the remaining thermodynamic quantities, we integrate the first law of thermodynamics $\mathrm dF=-S \,\mathrm dT$ to  get the free energy $F$ and the energy is then $M=F+TS$. We could, alternatively, read the energy directly at spatial infinity, which lies in the $1/r^6$ term in an asymptotic expansion. This would require that we accurately compute six derivatives at spatial infinity. While this is possible to do with enough precision, we opt to use the first law instead. We then tested the accuracy of the first law method by performing high resolution runs on a couple of points in moduli space, and compared the result for the energy extracted directly at spatial infinity with the results obtained from extracting the energy via the first law. The methods revealed an agreement that could be as small as $0.1\%$ for the high resolutions runs.  With the vacuum Einstein thermodynamics, $\{M,S,T,F\}$, the SYM thermodynamics $\{ \varepsilon, \sigma, \tau \}$ can be obtained from \eqref{map} and \eqref{dimlessthermo}. The free energy is then $\mathfrak{f}=\varepsilon-\tau\sigma$.

\subsection{Localised black holes} \label{sec:localized}
Within 10-dimensional vacuum Einstein gravity, localised black holes are asymptotically  ${\mathbb R}^{(1,8)}\times \widetilde{S}^1$ black holes with horizon topology $S^8$. For small energies, a perturbative construction of these solutions is available \cite{Harmark:2003yz,Gorbonos:2004uc,Harmark:2004ws,Dias:2007hg}, as reviewed in subsection \ref{ThermoUnifPert}. For higher energies, one must resort to numerical methods which we present here. The perturbative results will provide a valuable check on our numerics, while our numerical results will assess the regime of validity of the perturbative expansions \eqref{thermoLocGRpert}-\eqref{ThermoSYMpertLoc}.

Like nonuniform strings, localised black holes, are also static and axisymmetric.  However, the nonuniform strings contain a horizon that covers the entire axis, while the axis is partially exposed in localised black holes. This introduces a fifth boundary in the integration domain which complicates the construction of localised black holes. A suitable reference metric for localised black holes therefore must contain an axis, a topologically $S^8$ horizon, and asymptote to ${\mathbb R}^{(1,8)}\times \widetilde{S}^1$.  Furthermore, there is a periodic coordinate containing a $\mathbb Z_2$ symmetry, which we use to halve the integration domain.   To accommodate these five boundaries, we opt to work in two different coordinate systems. One of these is adapted to infinity, and the other to the horizon.

Let us now design our reference metric, beginning with the coordinates adapted to infinity. Our starting point is the ${\mathbb R}^{(1,8)}\times \widetilde{S}^1$ solution
\begin{equation}
\mathrm ds^2_{{\mathbb R}^{(1,8)}\times \widetilde{S}^1}=-\mathrm dT^2+\mathrm dR^2+R^2\mathrm d\Omega_7+d\widetilde{x}^2\;,
\end{equation}
where $\widetilde{x}\in(-\frac{\widetilde{L}}{2},\frac{\widetilde{L}}{2})$ is periodic. Scale out $\widetilde{L}$ by using the redefinitions $T=\pi\, t/\widetilde{L}$, $R=\pi \,r/\widetilde{L}$, and $\widetilde{\theta}=\pi \,\widetilde{x}/\widetilde{L}$ to get
\begin{equation}
\mathrm ds^2_{{\mathbb R}^{(1,8)}\times \widetilde{S}^1}=\frac{\widetilde L^2}{\pi^2}\bigg(-\mathrm dt^2+\mathrm dr^2+\mathrm d\widetilde{\theta}^{\,2}+r^2\mathrm d\Omega_7\bigg)\;,
\end{equation}
where $\widetilde{\theta}\in(-\pi/2,\pi/2)$ is periodic. Then perform a change of coordinates $r=\rho\sqrt{2-\rho^2}/(1-\rho^2)$ and $\widetilde{\theta}=2\arcsin(\xi/\sqrt 2)$ to
\begin{equation}
\mathrm ds^2_{{\mathbb R}^{(1,8)}\times \widetilde{S}^1}=\frac{\widetilde L^2}{\pi^2}\bigg[-\mathrm dt^2+\frac{4\mathrm d\rho^2}{(2-\rho^2)(1-\rho^2)^4}+\frac{4\mathrm d\xi^2}{2-\xi^2}+\frac{\rho^2(2-\rho^2)}{(1-\rho^2)^2}\mathrm d\Omega_7\bigg]\;.
\end{equation}
Now the coordinate ranges are the more convenient $\rho\in[0,1]$ and $\xi\in[-1,1]$.  Since we will be exploiting the $\mathbb Z_2$ symmetry in $\xi$, we will instead take $\xi\in[0,1]$ and demand reflection symmetry at $\xi=0$ and $\xi=1$. There is an axis at $\rho=0$ and asymptotic infinity is at $\rho=1$.  

At this point, we take our reference metric to be
\begin{equation}\label{reffar}
\overline{\mathrm ds}^2=\frac{\widetilde L^2}{\pi^2}\bigg[-m\,\mathrm dt^2+g\bigg(\frac{4\mathrm d\rho^2}{(2-\rho^2)(1-\rho^2)^4}+\frac{4\mathrm d\xi^2}{2-\xi^2}+\frac{\rho^2(2-\rho^2)}{(1-\rho^2)^2}\mathrm d\Omega_7\bigg)\bigg]\;,
\end{equation}
where $m$ and $g$ are functions of $\rho$ and $\xi$ that we need to specify.  We have chosen this particular form \eqref{reffar} for a number of reasons.  First, note that the $\mathrm d\rho^2$ and $\mathrm d\xi^2$ components in the metric came directly from $\mathrm dr^2+\mathrm d\widetilde{\theta}^{\,2}$, which is manifestly flat. It is therefore straightforward to transform \eqref{reffar} to other orthogonal coordinates. Our aim is to perform such a coordinate transformation where we can easily choose $m$ and $g$ so that the reference metric describes a black hole.  Second, when the black hole has small energy (high temperature), we would like the reference geometry near the horizon to resemble asymptotically flat 10-dimensional Schwarzschild-Tangherlini, as expected of the final solution. With the reference metric in the form \eqref{reffar}, it is easier to accommodate this by using Schwarzschild-Tangherlini in isotropic coordinates.

\begin{figure}[ht]
\centering
\includegraphics[width=.5\textwidth]{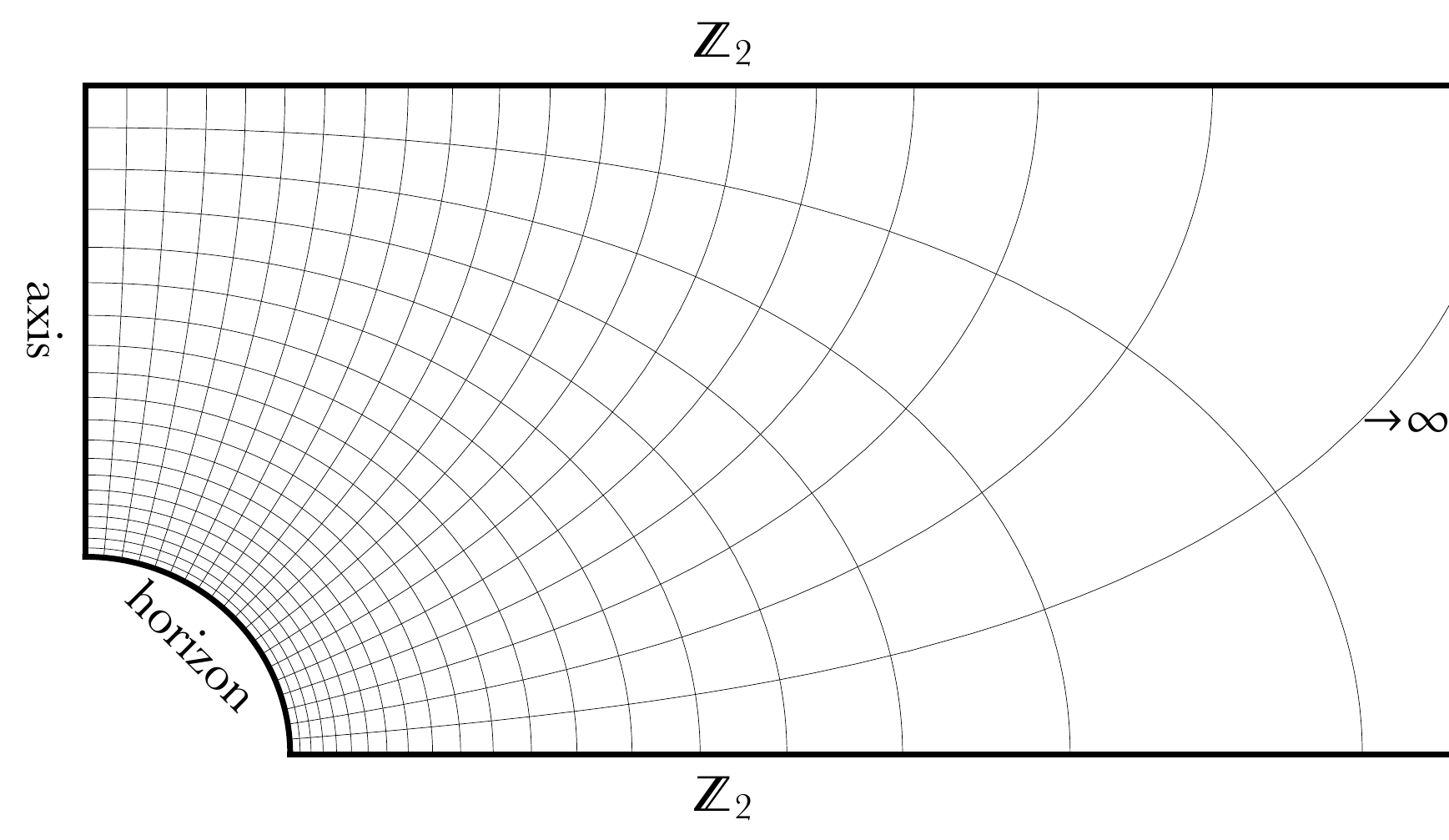}
\caption{Sketch of integration domain in $\{r,\widetilde\theta\}$ coordinates. The $\{\rho,\xi\}$ coordinates we use are directly related to these coordinates via $r=\rho\sqrt{2-\rho^2}/(1-\rho^2)$ and $\widetilde{\theta}=2\arcsin(\xi/\sqrt 2)$.  The grid lines are lines of constant $\chi$ and constant $y$.}\label{Fig:coord}
\end{figure}

The coordinate transformation we use and its inverse are given by
\begin{eqnarray}\label{coordmap}
   && \chi=\sqrt{1-\frac{\sinh {\bigl( }\frac{\rho  \sqrt{2-\rho ^2}}{1-\rho ^2}{\bigr) }}{\sqrt{\xi^2\left(2-\xi ^2\right)+\sinh^2{\bigl( }\frac{\rho  \sqrt{2-\rho^2}}{1-\rho ^2}{\bigr) }}}},  \qquad y=\frac{y_0\left(1-\xi ^2\right)}{\sqrt{\xi^2\left(2-\xi ^2\right)+\sinh^2{\bigl( }\frac{\rho  \sqrt{2-\rho^2}}{1-\rho ^2}{\bigr) }}} \,;\nonumber \\
   && \rho=\sqrt{1-\frac{1}{\sqrt{1+{\rm arcsinh}^2{\bigl( }\frac{y_0 \left(1-\chi^2\right)}{\sqrt{y^2+y_0^2  \, \chi^2 \left(2-\chi^2\right)}}{\bigr) }}}}, \qquad \xi=\sqrt{1-\frac{y}{\sqrt{y^2+y_0^2 \, \chi^2 \left(2-\chi^2\right)}}},
\end{eqnarray}
which can be derived from a conformal mapping.\footnote{It is also a mapping from elliptic coordinates to bipolar coordinates, both of which can be obtained from a conformal mapping of Cartesian coordinates.}  In these new coordinates, the reference metric becomes
\begin{align}\label{refnear}
      \overline{ds}^2= \frac{{\widetilde L}^2}{\pi^2} {\bigg \{} -m\,{\mathrm d}t^2+g {\biggl [} \frac{y_0^2}{h} \left( \frac{{\mathrm d}y^2}{y^2+y_0^2} +\frac{4\mathrm d\chi^2}{2-\chi^2}\right) +\, s \left(1-\chi^2\right)^2  {\mathrm d}\Omega_7^2 {\biggr ]}  {\bigg \}}\;,
\end{align}
where we assume $m$ and $g$ transform as scalars, $y_0>0$ and
\begin{equation}\label{hsdef}
 h=y^2 + y_0^2 \chi^2 \left(2-\chi^2\right) \,,\qquad s=\frac{{\rm arcsinh}^2{\bigl (}\frac{y_0 \left(1-\chi^2\right)}{\sqrt{y^2+y_0^2 \,\chi^2 \left(2-\chi^2\right)}}{\bigr )}}{\left(1-\chi^2\right)^2}.
\end{equation}
Note that $s$ is positive definite and regular, even at $\chi=1$. $h$ is positive except at $\chi=y=0$, where it vanishes. In these new coordinates, the axis is at $\chi=1$ and asymptotic infinity is at the coordinate `point' $\chi=y=0$.  The locations $\xi=0$ and $\xi=1$ where we require reflection symmetry have been mapped to $\chi=0$ and $y=0$, respectively. The location $\rho=0$, $\xi=0$ has been mapped to $y\to\infty$. We have introduced the constant $y_0$ in anticipation of placing a horizon at $y=1$.  $y_0$ therefore moves the location of this horizon in the $\{\rho, \xi\}$ coordinates. A small $y_0$ resembles a `small' black hole, and a large $y_0$ resembles a `large' black hole. A sketch of the integration domain and grid lines of constant $\chi$ and $y$ are shown in Fig. \ref{Fig:coord}.

Now we choose the functions $m$ and $g$.  These functions must satisfy a number of requirements. Our asymptotics requires that they must approach $1$ at $\chi=y=0$.  The reflection symmetries requires that $m$ and $g$ be even functions of $y$ and $\chi$. Lastly, they must give a regular horizon. Ideally, we would also like the geometry near the horizon for small $y_0$ to resemble asymptotically flat Schwarzschild-Tangherlini in isotropic coordinates, which in $d$-dimensions can be written
\begin{equation}\label{schwiso}
\mathrm ds^2_{\mathrm{Schw}}=-\left(\frac{1-y^{d-3}}{1+y^{d-3}}\right)^2\mathrm dt^2+y_0^2(1+y^{d-3})^{\frac{4}{d-3}}\left(\frac{\mathrm dy^2}{y^4}+\frac{1}{y^2}\mathrm d\Omega_{d-2}\right)\;.
\end{equation}
Note that these coordinates look similar to the $y_0\to0$ limit of our reference metric \eqref{refnear}. This suggests that we choose
\begin{equation}\label{mgdef}
m=\frac{1}{1+y_0^2\chi^2(2-\chi^2)}\left(\frac{1-y^6}{1+y^6}\right)^2\;,\qquad g=(1+y^6)^{2/3}\;.
\end{equation}
Note that we used $d=9$ components from \eqref{schwiso} rather than $d=10$ because of the requirement that $m$ and $g$ be even functions of $y$. The extra $\chi$ dependence in $m$ is placed to fix regularity of the horizon. These functions can be mapped back to $\{\rho,\xi\}$ coordinates through \eqref{coordmap}.  

To summarise, we now have a reference metric in two coordinate systems \eqref{reffar}, \eqref{refnear} where auxiliary functions given in \eqref{hsdef}, \eqref{mgdef}, and the coordinates are related by \eqref{coordmap}. We now give our metric ansatz:
\begin{align}\label{localisedansatz}
      {\mathrm ds}^2 &= \frac{{\widetilde L}^2}{\pi^2} {\biggl \{} -m \tilde{f}_1{\mathrm d}t^2+g {\biggl [} \frac{4 \tilde{f}_2\,{\mathrm d}\rho^2 }{(2-\rho^2)(1-\rho^2)^4}
+\frac{4\tilde{f}_3}{2-\xi^2}\left( d\xi -\tilde{f}_5 \frac{\xi(2-\xi^2)(1-\xi^2)\rho}{(1-\rho^2)^2}d\rho \right)^2
+\tilde{f}_4\frac{\rho^2(2-\rho^2)}{(1-\rho^2)^2} \, {\mathrm d}\Omega_7^2  {\biggr ]}    {\biggr \}},\nonumber\\
&= \frac{{\widetilde L}^2}{\pi^2} {\Biggl (} -f_1\,{\mathrm d}t^2+g {\biggl \{} \frac{y_0^2}{h} \bigg[ \frac{f_2\,{\mathrm d}y^2}{y^2+y_0^2} +\frac{4 f_3}{2-\chi^2} \bigg( {\mathrm d}\chi -f_5\,\frac{\chi \left(2-\chi^2\right) \left(1-\chi^2\right) y \left(1-y^2\right)}{h}\,{\mathrm d}y  \bigg)^2\bigg] \nonumber \\
      & \quad\quad\quad\quad\quad\quad\quad\quad\quad\quad +f_4\, s \left(1-\chi^2\right)^2  {\mathrm d}\Omega_7^2 {\biggr \}}  {\Biggr )}\;,
\end{align}
where $\tilde f_i$ are unknown functions of $\{\rho,\xi\}$, and $f_i$ are unknown functions of $\{\chi,y\}$. The known functions should instead be treated as scalars, transforming between coordinate systems as \eqref{coordmap}. 

Now we discuss boundary conditions. At the asymptotic boundary $\rho=1$ we impose that the solution approaches the reference metric, which is a Dirichlet condition. Regularity at the horizon $y=1$ requires that $f_i$ obey certain Robin boundary conditions (with expressions that are long and unilluminating). $f_i$ and $\tilde{f}_i$ obey Neumann conditions at the remaining boundaries either due to regularity at the axis $\rho=0$ (or $\chi=1$), or reflection symmetry at $\xi=0$ (or $\chi=0$) and at $\xi=1$ (or $y=0$).

\begin{figure}[hb]
\centering
\includegraphics[width=.35\textwidth]{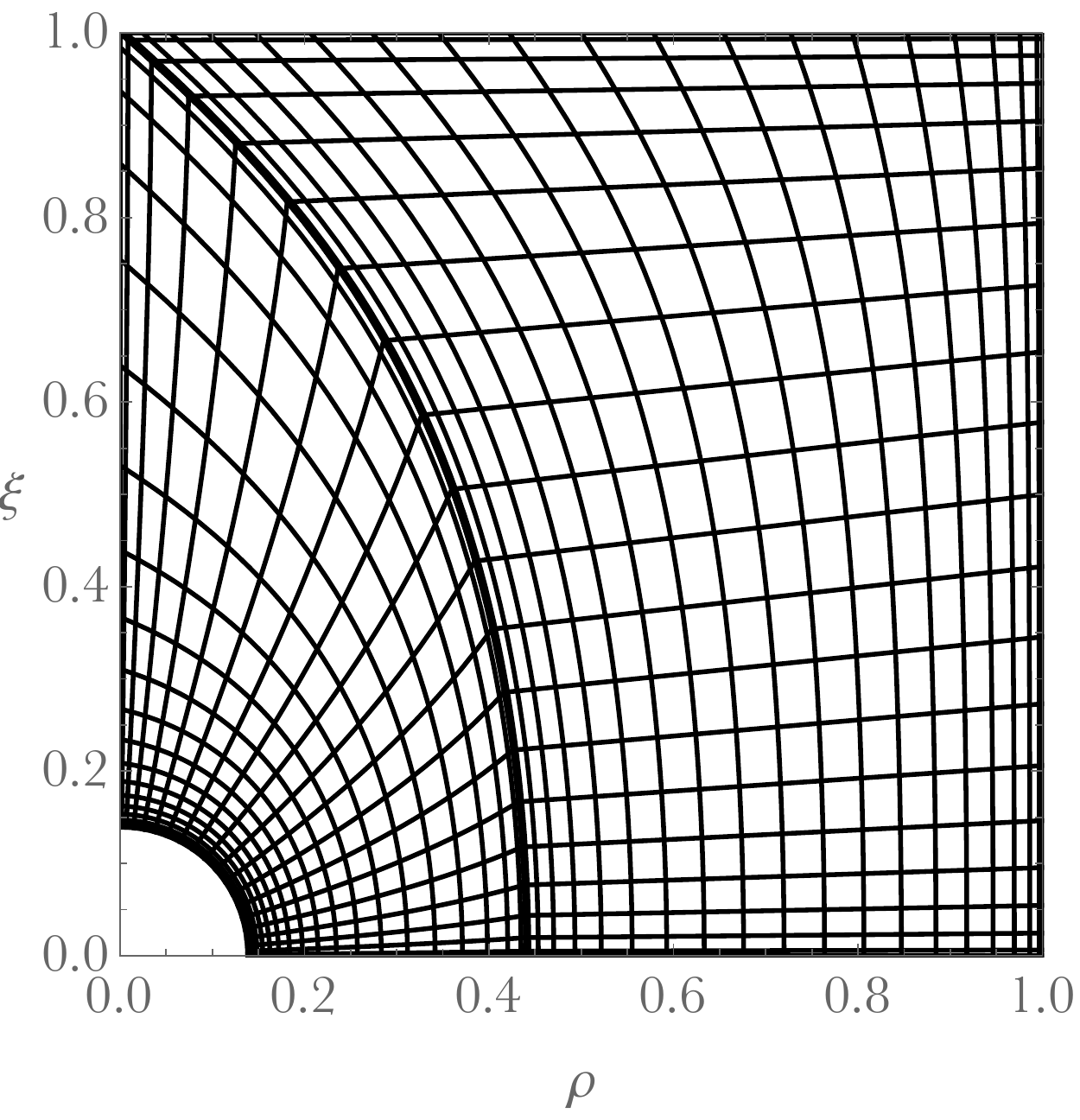}
\caption{Integration domain with two patches in $\{\rho,\xi\}$ coordinates.  Chebzshev-Gauss-Lobatoo grids are placed using transfinite interpolation. The patch near the horizon (the quarter circle in the lower left) is mapped from $\{\chi,y\}$ coordinates using \eqref{coordmap}.}\label{Fig:patches}
\end{figure}

As in the nonuniform case, we solve the equations of motion \eqref{EdeT} numerically with a Newton-Raphson algorithm.  To discretise,  we divide the integration domain into two patches as shown in Fig.~\ref{Fig:patches}.  Chebyshev-Gauss-Lobatto grids are placed in each patch using transfinite interpolation \cite{Dias:2015nua}. At the patch boundaries, we require equivalence of the line elements given by \eqref{localisedansatz}, as well as equivalence of the normal derivative across the patch boundary. As a first seed, we choose the reference metric with $y_0=1/10$.

Given $f_i(\chi,y)$, the entropy and temperature of the localised black holes are: 
\begin{equation}\label{thermoLocGRnum}
S=\frac{{\widetilde L}^8}{G_{10}} \int_0^1 \mathrm d\chi\,\frac{2^{8/3}\,y_0 \,{\rm arcsinh}^7{\bigl( }\frac{y_0 \left(1-\chi^2\right)}{\sqrt{1+y_0^2\,\chi^2 \left(2-\chi^2\right)}} {\bigr) }}{3 \pi ^4 \sqrt{\left(2-\chi^2\right) \left(1+y_0^2\,\chi^2 \left(2-\chi^2\right)\right)}} \,\sqrt{f_3(\chi,1)}\, f_4(\chi,1)^{7/2}\,,\qquad\qquad 
T=\frac{1}{{\widetilde L}}\frac{3}{2^{4/3}}\frac{\sqrt{1+y_0^2}}{y_0}.
\end{equation}
Our numerical solutions are parametrised by $y_0$, which defines the temperatures. $\widetilde L$ just sets a scale and drops out of our equations of motion. Note our choice of reference metric has restricted our temperature range to $T\widetilde L<3/2^{4/3}\approx 1.19$. We are actually not far from saturating this bound, but the purpose of this manuscript is not to study how localise black holes merge with non-uniform strings. Instead, what we want is to show a phase transition exists, and this turns out to occur for values of $T \widetilde L$ smaller than $3/2^{4/3}$. To get the free energy $F$ we integrate the first law of thermodynamics, $\mathrm dF=-S \,\mathrm dT$, and the energy is then $M=F+TS$.  The SYM thermodynamics $\{ \varepsilon, \sigma, \tau \}$ can be obtained via \eqref{map} and \eqref{dimlessthermo}. The free energy is $\mathfrak{f}=\varepsilon-\tau\sigma$.

\section{Thermodynamics of SYM on a circle at strong coupling.} \label{sec:Results}

In this section we present the thermodynamic phase diagrams of SYM on $\mathbb R^{(t)}\times S^1$ at large $N$ and strong coupling $\lambda^{\prime}$ as obtained from the gravitational side of the correspondence. The complementary phase diagrams within vacuum gravity can be found in Appendix \ref{sec:ResultsGR}. 

Recall that the thermodynamic quantities of the uniform SYM phase is given in \eqref{ThermoSYMunif} and low-energy perturbative expressions for the thermodynamic quantities of localised SYM phase are given in \eqref{thermoLocGRpert}. The nonuniform phase and the localised phase at higher energies are obtained numerically as described in the end of subsections \ref{sec:SetupNonUnif} and \ref{sec:localized}. The phase diagram can be obtained in either the microcanonical ensemble where energy is fixed and the phase with highest entropy is dominant, or in the canonical ensemble where temperature is fixed and the phase with lowest free energy is dominant. 

\begin{figure}[ht]
\centering
\includegraphics[width=.45\textwidth]{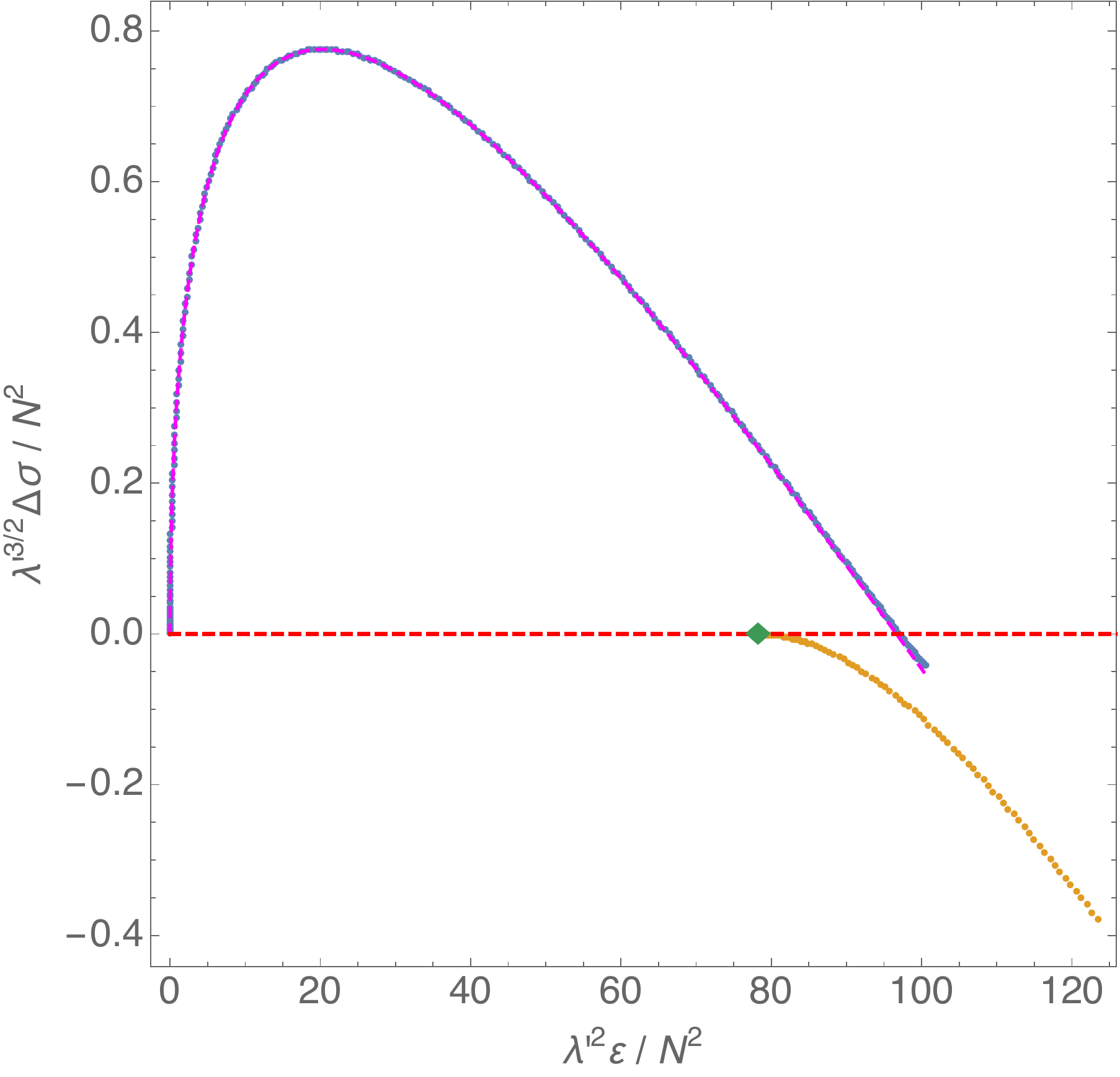}
\hspace{0.5cm}
\includegraphics[width=.45\textwidth]{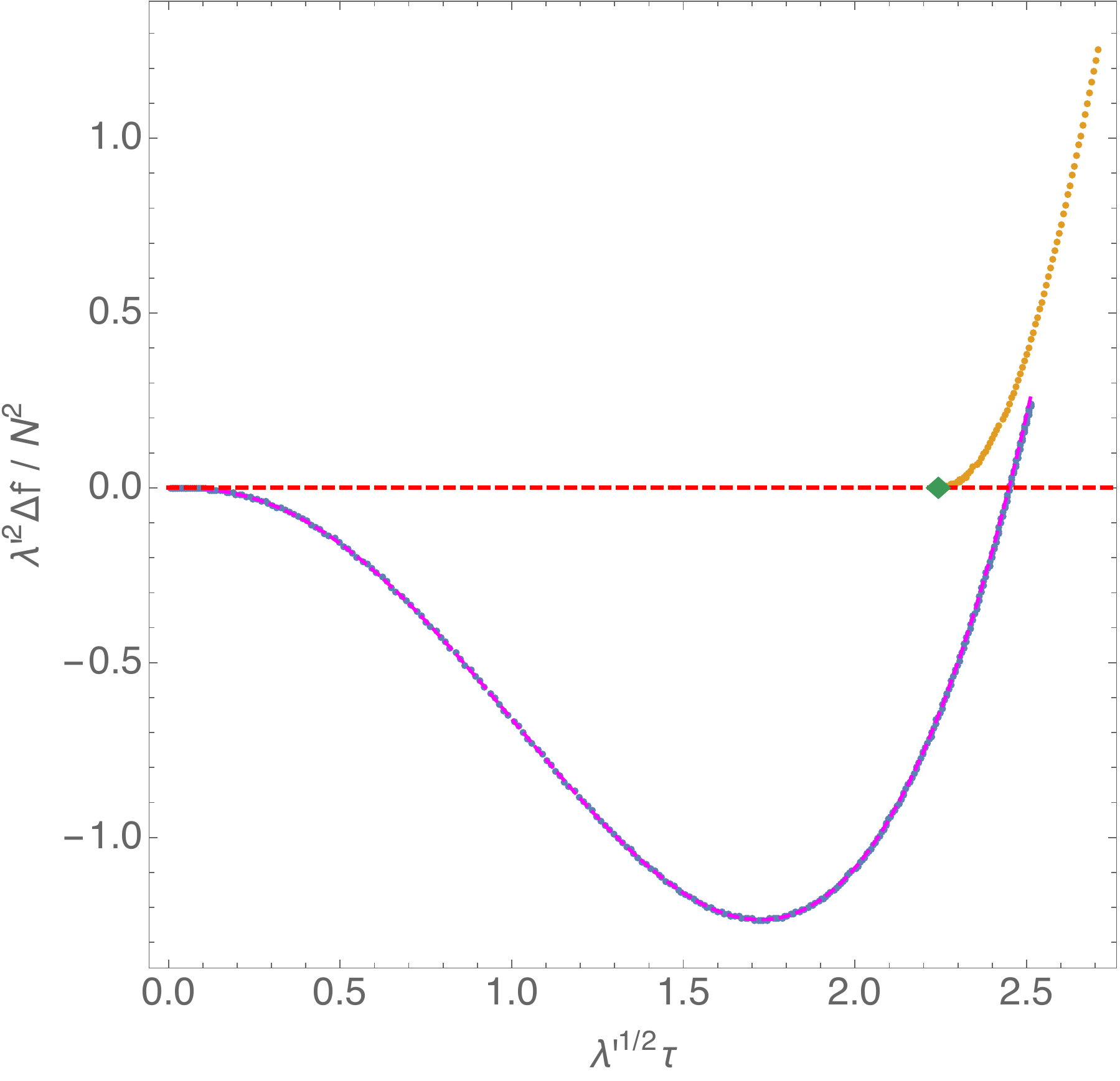}
\caption{{\bf Left panel:} Phase diagram of thermal phases of SYM$_{(1+1)}$ on a circle $S^1$ in the microcanonical ensemble.  $\Delta \sigma$ gives the entropy difference between a given thermal phase and the uniform phase with the same energy $\varepsilon$.  
{\bf Right panel:} Phase diagram of thermal phases of (1+1)-dimensional SYM on a circle $S^1$ in the canonical ensemble. $\Delta \mathfrak{f}$ gives the free energy difference between a given thermal phase and the uniform phase with the same energy $\varepsilon$. 
In these diagrams the horizontal (dashed red) line describes the uniform thermal phase. The orange points represents the nonuniform thermal family and the blue points describes the localised thermal phase.  The magenta dashed line represents the perturbative results for localised black holes.}\label{Fig:thermoSYM}
\end{figure}

The phase diagram in the microcanonical ensemble is presented in the left panel of Fig. \ref{Fig:thermoSYM}. We plot the dimensionless entropy difference, $\lambda^{\prime\,3/2}\Delta\sigma/N^2$, where $\Delta\sigma=\sigma(\varepsilon)-\sigma_0(\varepsilon)$, with $\sigma_0$ being the entropy of the uniform phase. The horizontal dashed line is the uniform thermal phase. It is unstable for $\varepsilon<\varepsilon_{GL}$ with the GL zero mode being at $\varepsilon_{GL}= 77.988 \, \frac{N^2}{{\lambda^\prime}^2}$ (green diamond in Fig. \ref{Fig:thermoSYM}). This GL zero mode is a bifurcation point to the orange curve branch that describes the nonuniform SYM phase. This nonuniform phase exists for $\varepsilon> \varepsilon_{GL}$ and $\Delta\sigma <0$. The uniform phase is therefore preferred over the nonuniform phase. 

The localised phase is represented by the blue curve. This curve intersects that of the uniform phase at an energy
\begin{equation}\label{MicroSYMphaseTr}
\varepsilon_{\rm PhT}= 97.067 \,\frac{N^2}{{\lambda^\prime}^2} = 1.245 \, \varepsilon_{\rm GL}\;,
\end{equation}
For low energies $\varepsilon<\varepsilon_{\rm PhT}$, the localised phase is the solution with highest entropy and thus the one that dominates the microcanonical ensemble.  However, the uniform phase becomes dominant for  $\varepsilon>\varepsilon_{\rm PhT}$. The phase transition is first order.

The phase diagram in the canonical ensemble is shown in the right panel of Fig. \ref{Fig:thermoSYM}, where we plot the dimensionless free energy difference, $\lambda^{\prime\,2}\Delta\mathfrak{f}/N^2$, where $\Delta\mathfrak f=\mathfrak f(\tau)-\mathfrak f_0(\tau)$, with $\mathfrak f_0$ being the free energy of the uniform phase. The colour scheme is the same as in the microcanonical ensemble. The GL zero mode is at $\tau_{GL}= 2.243 /\sqrt{\lambda^{\prime}}$ (green diamond) where the uniform thermal phase (horizontal dashed line) is unstable for $\tau<\tau_{GL}$.  We find that the localised phase has lowest free energy and is therefore dominant for $\tau<\tau_{\rm PhT}$, and the uniform phase is dominant for $\tau>\tau_{\rm PhT}$. The transition at
\begin{equation}\label{CanonicalSYMphaseTr}
\tau_{\rm PhT}=  \frac{2.451}{\sqrt{\lambda^{\prime}}}= 1.093 \, \tau_{\rm GL}   
\end{equation}
is first order.  Finally, the nonuniform phase is never dominant. 

As in studies of the same system in other dimensions \cite{Kol:2002xz,Kol:2003ja,Asnin:2006ip,Harmark:2002tr,Wiseman:2002ti,Kudoh:2003ki,Kudoh:2004hs,Sorkin:2006wp,Kleihaus:2006ee,Harmark:2007md,Headrick:2009pv,Wiseman:2011by,Figueras:2012xj,HorowitzBook2012,Kalisch:2016fkm}, we expect the uniform and localised phases to merge at a conical transition point.  To be consistent with the first law, one of these curves must form a cusp in the phase diagrams of Fig. \ref{Fig:thermoSYM}.  Close to the merger, it is expected that the phase diagram will develop an intricate zig-zagged line structure with a (possibly infinite) series of nearby cusps \cite{Kleihaus:2006ee,Kalisch:2016fkm}.  Unfortunately, we were unable to find this cusp, let alone the merger, with our present numerical methods and resources.

Now we compare our numerical results with the perturbative results \eqref{ThermoSYMpertLoc} for the localised phase.  Recall that the perturbative results are expected to be valid strictly for low energies, which in SYM thermodynamics also corresponds to low temperatures. However, we find that the perturbative results agree remarkably well with the numerical results all the way up (and even beyond) the first order phase transition point. This is best illustrated in Figs. \ref{Fig:thermoSYM}: the perturbative prediction is represented by the magenta dashed line and it essentially agrees with the numerical curve. This tests our numerical results and concludes that we can use the perturbative results \eqref{ThermoSYMpertLoc}  for the localized SYM phase well beyond what would be expected. This strong agreement may be due in part to the high powers within the perturbative expansion.

\begin{figure}[ht]
\centering
\includegraphics[width=.45\textwidth]{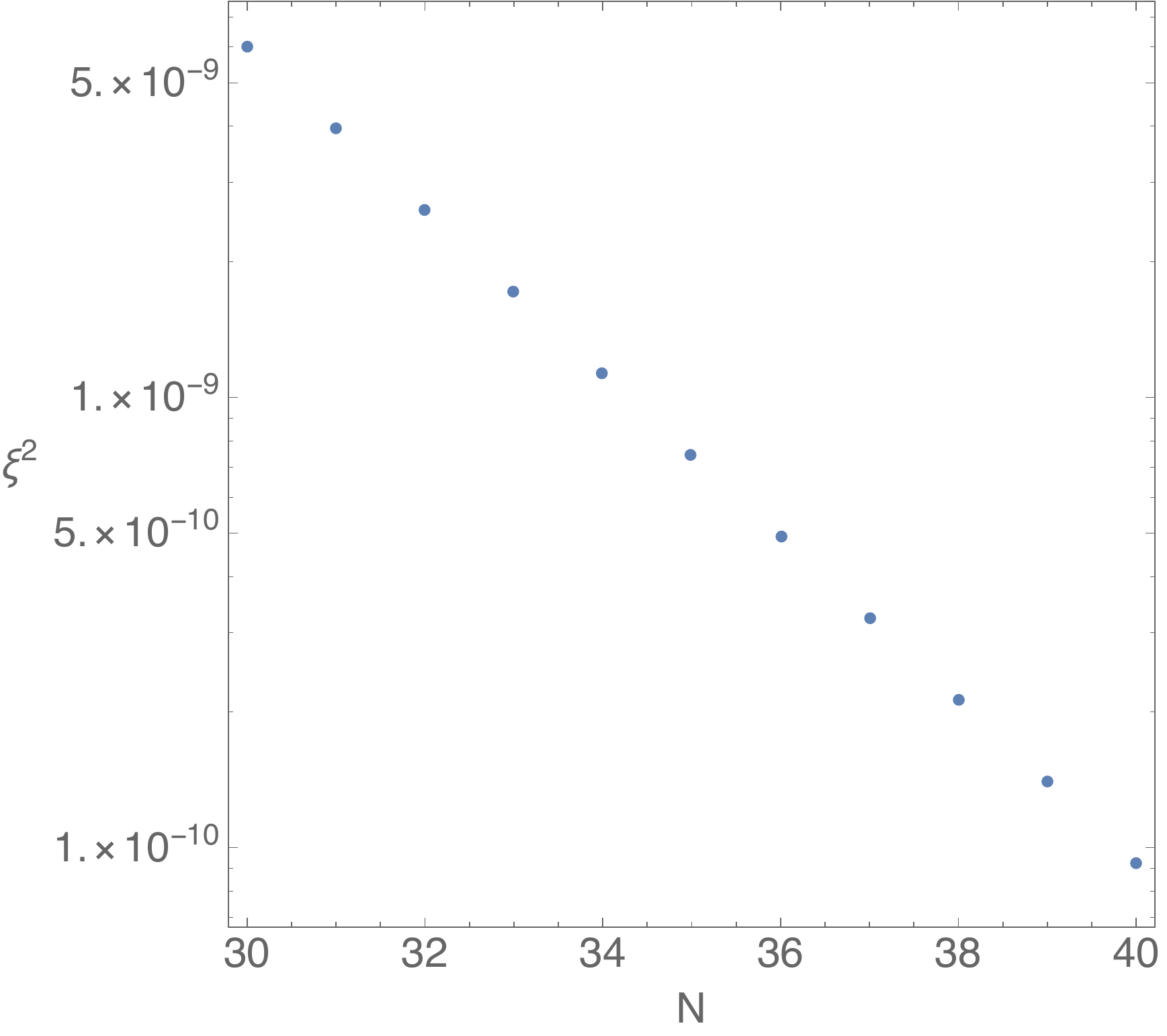}
\hspace{0.5cm}
\includegraphics[width=.45\textwidth]{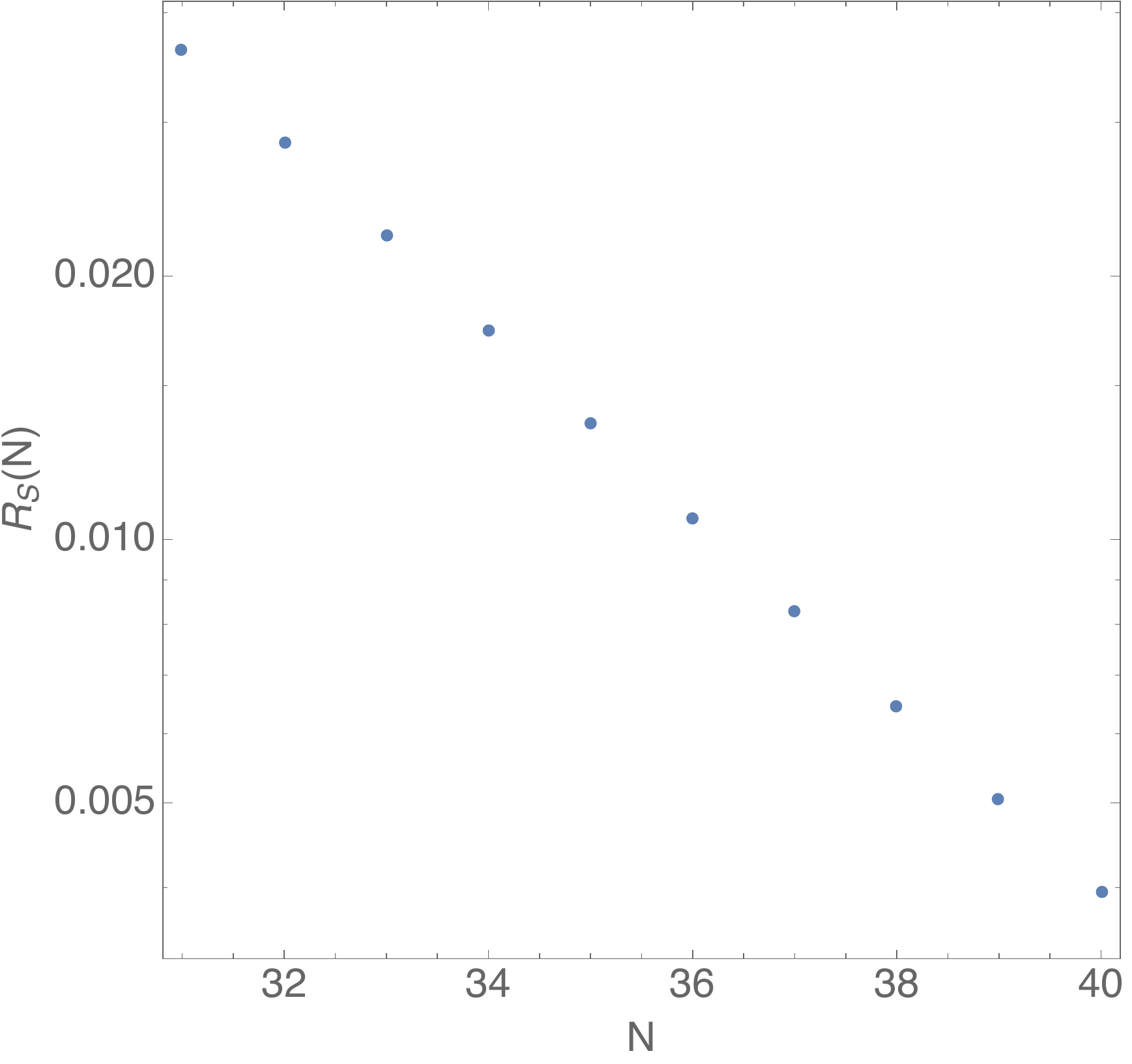}
\caption{Convergence of DeTurck norm (left) and entropy (right) for localised black holes at $y_0=5/2$.  Note the log scale indicating exponential convergence.}\label{Fig:convergence}
\end{figure}

Finally, let us demonstrate the convergence of our numerical method. In the left panel of Fig. \ref{Fig:convergence}, we plot the maximum value of the DeTurck norm $\xi^2$ for the localised black holes.  Recall that $\xi^\mu$ is guaranteed to be zero by the theorems in \cite{Figueras:2011va}. We find that this value vanishes exponentially with increase grid size, as expected of our pseudospectral methods. In all plots of phase diagrams, we use a resolution high enough so that all numerical solutions satisfy $\xi^2<10^{-10}$. In the right panel, of  Fig. \ref{Fig:convergence}, we show the convergence of the entropy  $R_S(N) = \left|1-S_{N}/S_{N-1}\right|$.  This quantity also decreases exponentially, as expected.

\section{Discussion} \label{sec:Conclusion}
Let us recapitulate our main results.  Thermal states of SYM on $\mathbb R^{(t)}\times S^1$ with circle length $L$ have three dimensionless parameters: $N$ from the gauge group, the dimensionless coupling $\lambda^\prime=g_{\mathrm YM}^2NL^2$, and some dimensionless thermodynamic quantity like the energy $\varepsilon\frac{{\lambda^\prime}^2}{N^2}$ or temperature $\tau\sqrt{\lambda^{\prime}}$, here expressed in units of the circle length $L$.  At large $N$, large $\lambda^\prime$ and temperatures $\tau\ll \lambda^{\prime-1/6}$, the theory is conjectured to be dual to classical type IIA supergravity, describing the near-horizon geometry of a collection of $D0$ branes on a circle.  Through an uplift-boost-KK reduction procedure, solutions to classical type IIA supergravity can be generated from solutions of asymptotically $\mathbb R^{(1,8)}\times S^1$ vacuum Einstein solutions. We numerically construct nonuniform and localised phases within vacuum Einstein gravity and then map their thermodynamics to the conjectured SYM thermodynamic variables.  In the microcanonical ensemble, we find a first-order phase transition at
\begin{equation}\label{MicroSYMphaseTr2}
\varepsilon_{\rm PhT}\frac{{\lambda^\prime}^2}{N^2}= 97.067\;,
\end{equation}
while in the canonical ensemble we find a first-order phase transition at the temperature
\begin{equation}\label{CanonicalSYMphaseTr2}
\tau_{\rm PhT}\sqrt{\lambda^{\prime}}= 2.451\;.
\end{equation}
The phase transition is between a uniform phase at higher energies and temperatures to a localised phase at lower energies and temperatures.  From the perspective of the gauge theory, this phase transition represents spontaneous symmetry breaking of the $U(1)$ symmetry of the $S^1$ \cite{Aharony:2004ig,Aharony:2005ew,Catterall:2010fx}.

These results can be compared with available lattice and perturbative results on the gauge theory side \cite{Aharony:2004ig,Azeyanagi:2009zf,Hanada:2009hq,Hanada:2010kt,Hanada:2010gs,Catterall:2010fx,Hanada:2016qbz} performed at large $N$, which we will now summarise.  At weak coupling, perturbative SYM indicates that the theory undergoes a second order confinement/deconfinement phase transition (or center symmetry breaking) at $\tau\,\lambda^\prime \gtrsim 1.35$ \cite{Aharony:2004ig,Mandal:2009vz,Kawahara:2007ib}. As we have mentioned, at strong coupling, the gravity side conjectures a first order phase transition. Due to strong coupling, this phase transition is nontrivial from the perspective of the gauge theory.  Nevertheless, such a phase transition can be sought through lattice simulations.  Fortunately, some lattice calculations for SYM on $\mathbb R^{(t)}\times S^1$ are already available \cite{Catterall:2010fx}\footnote{See also the recent lattice study \cite{Hanada:2016qbz} on the bosonic SU(N) YM theory which upgrades to to SYM when fermions with anti-periodic boundary conditions on $S^1$ are added}  and ongoing work is improving these simulations \cite{Catterall:2017}. References \cite{Aharony:2004ig,Aharony:2005ew} conjectured that the weak coupling confinement/deconfinement and strong coupling uniform/localised phase transitions should be a continuation of each other in a phase diagram that scans all the coupling values. Available lattice computations are consistent with this picture. In lattice simulations, good observables for the confinement/deconfinement and uniform/localised phase transitions are the expectation value of the trace of Polyakov or Wilson loops along the Euclidean time and spatial circles, as well as the associated distribution of eigenvalues of the Polyakov/Wilson loops on these circles \cite{Gross:1980he,Susskind:1997dr,Barbon:1998cr,Li:1998jy,Fidkowski:2004fc,Aharony:2004ig,Aharony:2005ew,Catterall:2010fx,Hanada:2016qbz}. At strong coupling, this distribution of eigenvalues describes the positions of the collection of D0-branes in the type IIA system.  For example, the Wilson loop along the spatial circle yields a uniform distribution of eigenvalues, a nonuniform distribution without gap, or a localised distribution with a gap that clearly distinguish the uniform, nonuniform and localised SYM phases \cite{Aharony:2004ig,Aharony:2005ew,Catterall:2010fx,Hanada:2016qbz}.

The available lattice computations \cite{Catterall:2010fx,Hanada:2016qbz} are consistent with our critical values for the phase transition but an accurate confirmation will require an improved code that is, in particular, able to do the computations at higher $N$. In the near future, results from improved lattice simulations can be more accurately compared with our gravitational predictions of Fig. \ref{Fig:thermoSYM} and the critical values for the location of the phase transition \eqref{MicroSYMphaseTr2} and \eqref{CanonicalSYMphaseTr2}. In tandem, these results would comprise the first stringent tests of gauge/gravity duality involving a first-order phase transition.

\begin{acknowledgments}

Some computations were performed on the COSMOS Shared Memory system at DAMTP, University of Cambridge operated on behalf of the STFC DiRAC HPC Facility and funded by BIS National E-infrastructure capital grant ST/J005673/1 and STFC grants ST/H008586/1, ST/K00333X/1.  O.J.C.D. is supported by the STFC Ernest Rutherford grants ST/K005391/1 and ST/M004147/1. This research received funding from the European Research Council under the European Community's 7th Framework Programme (FP7/2007-2013)/ERC grant agreement no. [247252]. BW is supported by NSERC.
\end{acknowledgments}


\appendix
\section{Equations of motion for type II supergravity}
\label{sec:EOM}
For completeness, in this appendix we give the equations of motion that are obeyed by p-brane solutions of type II supergravity and their near-horizon geometries. This includes (non-)uniform and localized solutions.

In the string frame, the equations of motion of type II action \eqref{action} are
\begin{eqnarray}\label{IIeomStringFr}
&& R_{ab}=-2\nabla_a\nabla_b \phi +\frac{1}{4}  \frac{1}{(p+2)!}\,e^{2\phi}{\biggl [} 2(p+2)\, F_a^{\,\: c_1\cdots c_{p+1}}F_{b c_1\cdots c_{p+1}} - g_{ab} \,F_{c_1\cdots c_{p+2}}F^{c_1\cdots c_{p+2}}{\biggr ]}, \nonumber \\
&& \nabla_c F^{c a_1 \cdots a_{p+1}}=0\,,\nonumber \\
&& \nabla_c\nabla^c \phi- 2\partial_c \phi \partial^c \phi +\frac{1}{4} \frac{p-3}{(p+2)!}\, e^{2\phi}  F_{c_1\cdots c_{p+2}}F^{c_1\cdots c_{p+2}}=0\,,
\end{eqnarray}
where we have introduced the RR field strength $F_{(p+2)}=\mathrm dA_{(p+1)}$.

In the Einstein frame, type II action \eqref{action} reads
\begin{equation}
\label{actionE}
I_{II}^{(E)} = \frac{1}{16\pi G_{10}} \int \mathrm{d}^{10} x \sqrt{-\widetilde{g} } \, \Big( \widetilde{R}
- \frac{1}{2} \partial_\mu \phi \partial^\mu \phi - \frac{1}{2 (p+2)!} g_s^{\frac{1}{2}(p+1)} e^{\frac{1}{2}(3-p)\phi} (\mathrm dA_{(p+1)})^2 \Big),
\end{equation}
where Newton's constant is expressed in terms of the string length $\ell_s$ and string coupling $g_s$ as $16 \pi  G_{10}\equiv (2 \pi )^7 \ell_s^8  \,g_s^2$.
From this action we obtain the equations of motion
\begin{eqnarray}\label{IIeom}
&& \widetilde{R}_{ab}-\frac{1}{2}\,\widetilde{R} \,\widetilde{g}_{ab}=\frac{1}{2}\left(  \partial_a \phi \partial_b \phi - \frac{1}{2}\, \widetilde{g}_{ab}\,\partial_c \phi \partial^c \phi \right) \nonumber \\
&& \hspace{2.55cm} +\frac{1}{2}\,g_s^{\frac{1}{2}(p+1)} e^{\frac{1}{2}(3-p)\phi}\left( \frac{1}{(p+1)!}\, F_a^{\,\: c_1\cdots c_{p+1}}F_{b c_1\cdots c_{p+1}} -  \frac{1}{2 (p+2)!} \,\widetilde{g}_{ab}\,F_{c_1\cdots c_{p+2}}F^{c_1\cdots c_{p+2}}\right), \nonumber \\
&& \widetilde{\nabla}_c \left( e^{\frac{1}{2}(3-p)\phi} F^{c a_1 \cdots a_{p+1}}\right)=0\,,\nonumber \\
&& \widetilde{\nabla}_c\widetilde{\nabla}^c \phi- \frac{3-p}{4(p+2)!}\, g_s^{\frac{1}{2}(p+1)} e^{\frac{1}{2}(3-p)\phi}  F_{c_1\cdots c_{p+2}}F^{c_1\cdots c_{p+2}}=0\,,
\end{eqnarray}
where $\widetilde{\nabla}$ is the Levi-Civita connection of $\tilde{g}$.

Finally, recall that the string and Einstein frame metrics are related by the transformations $g_{ab}=\widetilde{g}_{ab} e^{\frac{1}{2}\left(\phi-\phi _{\infty}\right)}$, while the dilaton and gauge field are the same in both frames.
 In the main text, we present the type II solutions \eqref{nonExtDp}, \eqref{NHnonExtDp}, \eqref{NHnonExtD1}, \eqref{NHnonExtD0}, \eqref{soln:10KKUnif} and \eqref{FullnonExtD0} in the string frame.

\section{Thermodynamics in Vacuum Einstein Gravity}
\label{sec:ResultsGR}

Here we give the phase diagram of uniform, nonuniform, and localised solutions within asymptotically ${\mathbb R}^{(1,8)}\times \widetilde{S}^1$ vacuum Einstein gravity. These are the solutions we constructed explicitly in Section \ref{sec:NumConstruction} and whose thermodynamic quantities were used to get the SYM thermodynamics via the map \eqref{map}-\eqref{dimlessthermo}. This is the first time that the thermodynamics of the nonuniform and localised solutions associated with the GL instability are presented in 10 dimensions. The results of this appendix thus complement the studies done for other dimensions\cite{Gregory:1993vy,Gregory:1994bj,Kol:2002xz,Kol:2003ja,Harmark:2003yz,Gorbonos:2004uc,Harmark:2004ws,Asnin:2006ip,Harmark:2002tr,Wiseman:2002ti,Kudoh:2003ki,Kudoh:2004hs,Sorkin:2006wp,Kleihaus:2006ee,Harmark:2007md,Dias:2007hg,Headrick:2009pv,Wiseman:2011by,Figueras:2012xj,HorowitzBook2012,Kalisch:2016fkm}.

\begin{figure}[ht]
\centering
\includegraphics[width=.45\textwidth]{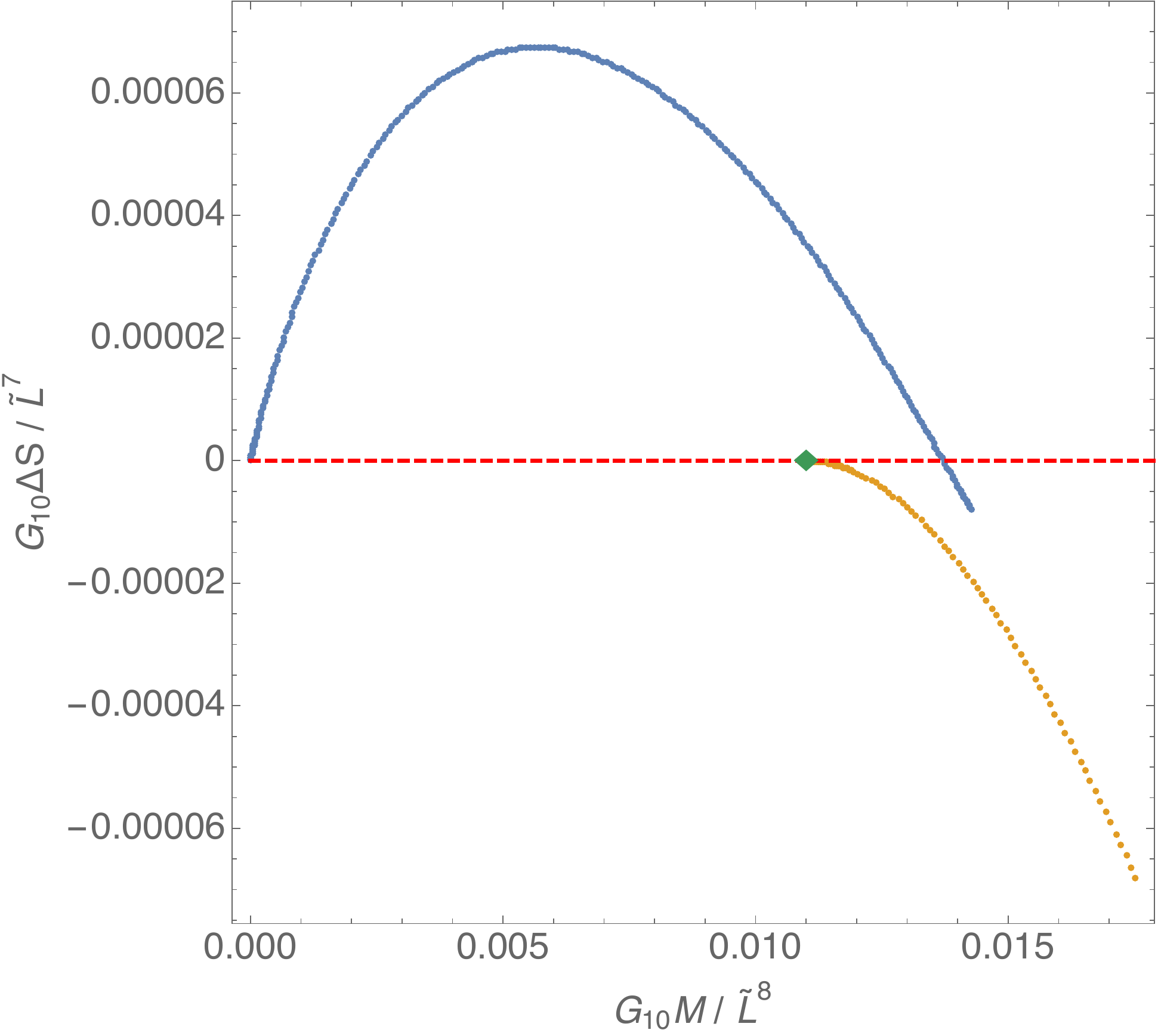}
\hspace{0.5cm}
\includegraphics[width=.45\textwidth]{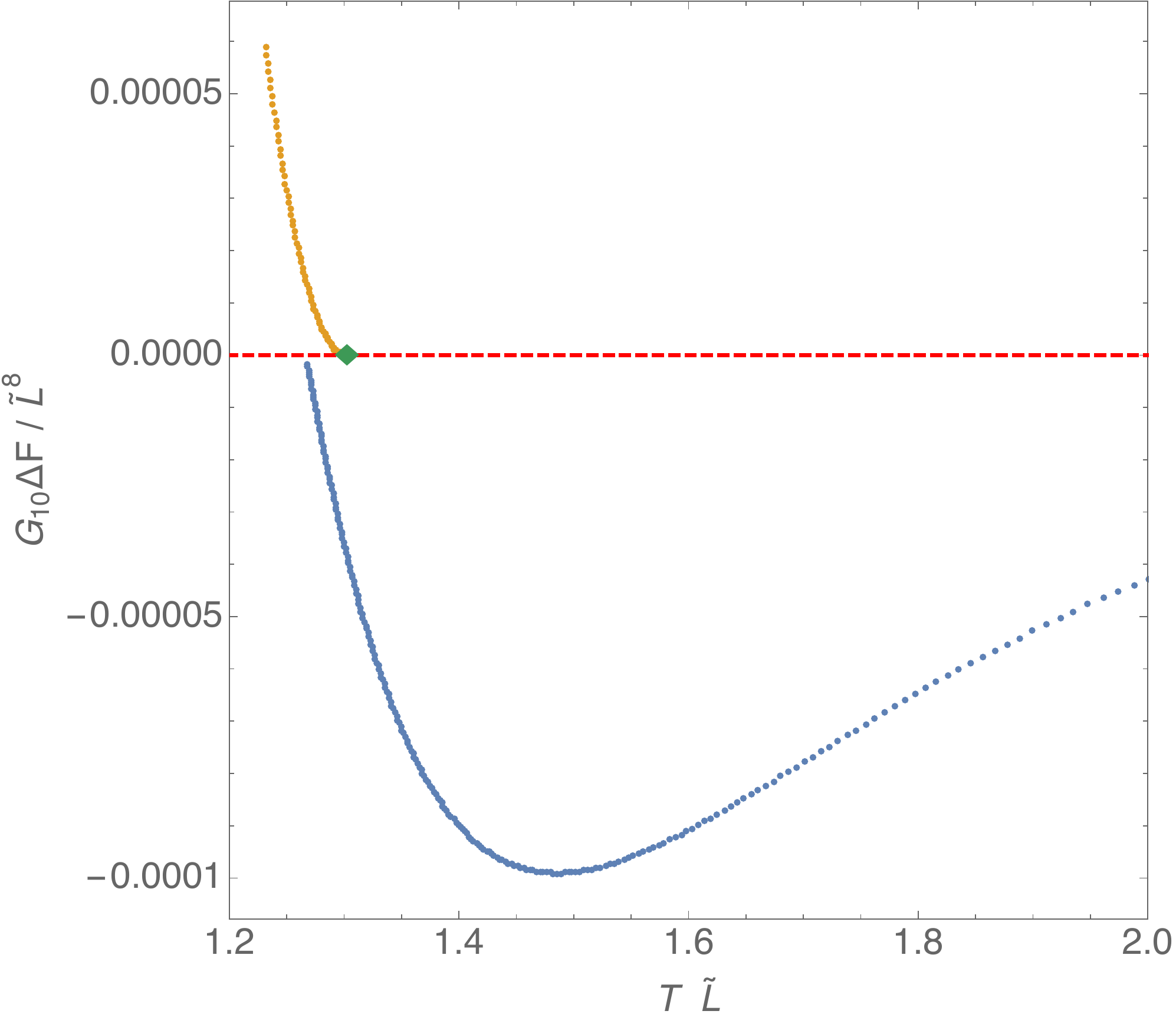}
\caption{{\bf Left panel:} Phase diagram of asymptotically ${\mathbb R}^{(1,8)}\times \widetilde{S}^1$ vacuum Einstein solutions in the microcanonical ensemble. $\Delta S$ gives the entropy difference between a given solution and the uniform black string with the same energy $E$.  
{\bf Right panel:} Phase diagram of asymptotically ${\mathbb R}^{(1,8)}\times \widetilde{S}^1$ vacuum Einstein solutions in the canonical ensemble. $\Delta F$ gives the free energy difference between a given solution and the uniform black string with the same temperature $T$.  
In these diagrams the horizontal (dashed red) line describes the uniform black string. The orange curve represents the nonuniform black string family and the blue curve describes the localised black hole branch.}\label{Fig:thermoGR}
\end{figure}

The phase diagram in the microcanonical ensemble is in the left panel of Fig. \ref{Fig:thermoGR}. We plot the dimensionless entropy difference $G_{10}\Delta S/\widetilde{L}^7$, between a given solution and the uniform black string as a function of energy $G_{10} M/\widetilde{L}^8$. Thus, the horizontal dashed line with $\Delta S=0$ describes the uniform black string family. This solution is unstable for energies below the GL zero mode at $M=M_{GL}$ with $M_{GL}=0.0110\,\widetilde{L}^8/G_{10}$ (labeled by the green diamond). The nonuniform black strings (orange curve) branch from the GL zero mode towards $M>M_{GL}$ and have less entropy than the uniform black string. The localised black hole (blue curve) dominate this ensemble for $M<M_{\rm PhT}$ At $M=M_{\rm PhT}$  with
\begin{equation}
M_{\rm PhT}= 0.0137 \,\frac{\widetilde{L}^8}{G_{10}}= 1.245 \,M_{GL}
\end{equation}
there is a first order phase transition with uniform strings being preferred for $M>M_{\rm PhT}$. 

Consider now the canonical ensemble. The phase diagram in this ensemble is displayed in the right panel of Fig. \ref{Fig:thermoGR} where we plot the dimensionless free energy difference $G_{10}\Delta F/\widetilde{L}^8$, between a given solution and the uniform black string, as a function of the temperature $T \widetilde{L}$.  The plot and colour scheme are the same as that of the microcanonical ensemble. The uniform black string (horizontal dashed line) is unstable for $T>T_{GL}$ with  $T_{GL}=1.302/\widetilde{L}$. There is a first order phase transition at
\begin{equation}
T_{\rm PhT}= \frac{1.266}{\widetilde{L}} = 0.972 \,T_{GL}
\end{equation}
where the uniform phase is preferred for lower temperatures and the localised phase is preferred for higher $T$. The nonuniform phase is never preferred. Note that our estimate for $T_{\rm PhT}$ involves a very small extrapolation, since in Fig.~(\ref{Fig:thermoGR}) we can observe that the free energy difference between the localised black hole phase and the uniform black string with the same temperature $T$, which we coined $\Delta F$, has not yet crossed zero.

\bibliography{refsD0}{}
\bibliographystyle{JHEP}

\end{document}